\documentclass[twocolumn]{autart}    

\usepackage{graphicx}          

\usepackage{amsmath,amssymb,amsfonts}
\usepackage{multirow}
\usepackage{siunitx}
\usepackage[shortlabels]{enumitem}
\usepackage{caption}
\usepackage{subcaption}

\newtheorem{assm}{Assumption}
\newtheorem{prb}{Problem}
\newtheorem{rmark}{Remark}
\newtheorem{lma}{Lemma}

\newtheorem{dfn}{Definition}

\newlist{properties}{enumerate}{2}
\setlist[properties]{label= P\arabic*:, font=\textbf, itemindent=*}

\usepackage{xcolor}

\DeclareMathOperator*{\proj}{proj}

\definecolor{hcolor}{rgb}{0,0,0}
\begin{document}

\begin{frontmatter}
	
	\title{Lagrangian-based online safe reinforcement learning for state-constrained systems} 
	
	\thanks[footnoteinfo]{This paper was not presented at any IFAC meeting.
		Corresponding author Soutrik Bandyopadhyay.}
	
	\author[IITD]{Soutrik
		Bandyopadhyay}\ead{Soutrik.Bandyopadhyay@ee.iitd.ac.in}, 
	\author[IITD]{Shubhendu Bhasin}\ead{sbhasin@ee.iitd.ac.in} 
	
	\address[IITD]{Department of Electrical Engineering, Indian Institute of
		Technology New Delhi} 

	\begin{keyword}
		Approximate dynamic programming;
		Safe reinforcement learning;
		Optimal control;
		Barrier Lyapunov functions.
	\end{keyword} 

	\begin{abstract}
		This paper proposes a safe reinforcement learning (RL) algorithm that approximately solves the state-constrained optimal control problem for continuous-time uncertain nonlinear systems. We formulate the safe RL problem as the minimization of a Lagrangian that includes the cost functional and a user-defined barrier Lyapunov function (BLF) encoding the state constraints. We show that the analytical solution obtained by the application of Karush-Kuhn-Tucker (KKT) conditions contains a state-dependent expression for the Lagrange multiplier, which is a function of uncertain terms in the system dynamics. We argue that a naive estimation of the Lagrange multiplier may lead to safety constraint violations. To obviate this challenge, we propose an Actor-Critic-Identifier-Lagrangian (ACIL) algorithm that learns optimal control policies from online data without compromising safety. We provide safety and boundedness guarantees with the proposed algorithm and compare its performance with existing offline/online RL methods via a simulation study.
	\end{abstract}
	
\end{frontmatter}

\section{Introduction}
\label{sec:intro}
Reinforcement learning (RL) has recently been used in control theory to solve
optimal control problems under uncertainties
\cite{bhatnagar2009Automatica,sutton2018reinforcement,al2008TSMC,vamvoudakis2010Automatica,bhasin2013Automatica,pang2022TAC}.
Leveraging the general framework
of RL, control-theoretic strategies have found reasonable success for
discrete-time \cite{bhatnagar2009Automatica,sutton2018reinforcement,al2008TSMC}
and continuous-time
\cite{vamvoudakis2010Automatica,bhasin2013Automatica,pang2022TAC} systems under
both deterministic and stochastic settings. Despite its successes, the
implementation of such learning-based controllers on safety-critical
systems is still challenging, primarily due to lack of provable safety
guarantees with on-policy RL algorithms. Specifically, due to inadequate
knowledge of the system dynamics during training, on-policy RL algorithms tend
to naively explore the state space without considering safety objectives. Thus,
solely relying on on-policy RL-based controllers may lead to safety violations.
To alleviate this challenge, the field of safe reinforcement learning (safe RL)
emerged, and researchers are actively seeking to bolster RL algorithms with
safety guarantees \cite{ames2016TAC,ames2019ECC,koller2018CDC,zanon2020TAC,berkenkamp2017NIPS,cohen2021TL,alshiekh2018safe,fisac2018general,cheng2019AAAI}.

In the context of safe RL-based control, ``safety'' often refers
to the satisfaction of user-defined constraints during training and deployment.
Specifically, safety formulations in dynamical systems often seek to ensure
invariance of constraint sets on state or control action
\cite{blanchini1999Automatica}. Based on this mathematical formulation, some
common approaches to ensuring safety for RL-based control include - using
control barrier functions (CBF) \cite{ames2016TAC,ames2019ECC}, model predictive
control (MPC) \cite{koller2018CDC,zanon2020TAC,berkenkamp2017NIPS}, temporal
logic (TL) specifications \cite{cohen2021TL,alshiekh2018safe}, and using
Gaussian processes \cite{fisac2018general,cheng2019AAAI}, to name a few. Control
barrier functions and its Lyapunov-like counterpart - barrier Lyapunov functions
(BLF) \cite{tee2009Automatica} provide a method for studying the forward
invariance of constraint sets through a Lyapunov-like analysis, without the
necessity to compute the system trajectories.

Another school of thought in solving optimal control problems under constraints
is to use numerical approaches (see e.g. \cite{hager2018SIAM,hager2019COA}
and \cite{rao2009survey} for a detailed survey) that provide
close-to optimal solutions for the desired objective functional under
user-defined state and actuation constraints. However, these methods usually
assume complete knowledge of the system and are generally computed offline.

To enable adaptation to parametric uncertainties online, several
on-policy RL algorithms have been proposed in the literature
\cite{vamvoudakis2010Automatica,bhasin2013Automatica,vamvoudakis2017SysConLet,kamalapurkar2016Automatica}
aiming to solve the optimal control problem in a continuous-time deterministic
setting. Inspired by these methods, numerous safe RL algorithms have been
proposed, such as barrier transformation based techniques
\cite{yang2019ACC,mahmud2021ACC,greene2020LCSS} which transform the constrained state into an
unconstrained state and subsequently apply approximate dynamic programming
techniques.
Another class of methods augment the barrier Lyapunov functions to
the cost functional of the optimal control problem
\cite{marvi2021IJRNC,cohen2020CDC}. While the barrier transformation methods can
only be applied for box constraints on the state, the methods in
\cite{marvi2021IJRNC,cohen2020CDC} rely on the assumption that the resulting
value function is continuously differentiable (see \cite{mahmud2021ACC}) which
may not hold for systems and safe sets in general.
The authors of \cite{cohen2023Automatica} proposed an online RL algorithm using
a ``safeguarding controller'' that uses the gradient of a
user-defined barrier Lyapunov function to modify the optimal control law in a
minimally invasive fashion to ensure safety. The method uses a
gain term to trade-off the safety and stability objectives.

\textcolor{hcolor}{
	Furthermore, recent developments in the field of safe RL
	for control include
	\cite{bandyopadhyay2023CDC,isaly2021ACC,kokolakis2022TNNLS,peng2023TSMC,cohen2023LCSS}.
	The authors of \cite{bandyopadhyay2023CDC} develop a safe Q-learning
	algorithm for continuous-time linear time invariant systems using reciprocal
	control barrier functions. In \cite{isaly2021ACC}, the authors use the integral
	concurrent learning technique along with multiple barrier functions to
	ensure safety under parametric system uncertainties.
	In \cite{kokolakis2022TNNLS}, a safety aware critic-only RL algorithm is developed that solves
	the persuit-evasion zero-sum differential game under uncertain environments.
	The authors of \cite{peng2023TSMC} develop a safe actor-critic algorithm to
	guide a missile to its target while avoiding unsafe areas. In
	\cite{cohen2023LCSS}, a general characterization of smooth safety filters
	is developed based on the Implicit function theorem.
}

A promising approach for computing safe RL policies is to
minimize the desired cost functional w.r.t. (with respect to) a constraint on
the time derivative of a user-defined candidate BLF or a candidate CBF. The
optimal control law for this formulation may be found using the
Karush-Kuhn-Tucker (KKT) conditions \cite{almubarak2021CDC} along with an
optimal Lagrange multiplier dependent on the state. However,
the optimal Lagrange multiplier includes unknown terms of the uncertain drift dynamics
and optimal value function. While
the authors of \cite{almubarak2021CDC} proposed an offline method using a safe Galerkin
successive approximation to learn the optimal control policy, we show that a
naive extension to an online algorithm fails to ensure the safety of the
underlying system under some non-trivial scenarios.

To alleviate this challenge, in this paper, we propose an on-policy RL
algorithm, utilizing an online estimate for the optimal Lagrange multiplier,
that approximately solves the optimal control problem online for a class of
deterministic continuous-time nonlinear systems with parametric uncertainties
under user-defined state constraints. The proposed
Actor-Critic-Identifier-Lagrangian (ACIL) algorithm extends the approach in
\cite{bhasin2013Automatica,kamalapurkar2016Automatica} to the constrained setting by including an online estimate of the Lagrange multiplier.
The proposed method is similar in spirit to \cite{cohen2023Automatica}; however, differs in the sense that while the gain for the
safeguarding controller in \cite{cohen2023Automatica} is held constant,
our approach uses an estimate of
the
Lagrange multiplier that varies according to the safety requirements of
the current state.
We rigorously show that the proposed approach guarantees both safety and stability.
Through simulations, we observe better performance in terms of
the total cost accrued.
Consequently, we argue that the use of the proposed Lagrange multiplier estimate
in an optimal controller leads to both safety enhancement and performance improvement.
\subsection{Contributions}

\label{sec:contrib}

\textcolor{hcolor}{
	In the literature, it is often observed that ensuring safety of uncertain
	systems often comes at the cost of optimality. Specifically, under
	parametric uncertainties of the system dynamics, safe RL algorithms
	prioritize safety and tend to be suboptimal w.r.t. the desired cost
	functional. This paper aims to bridge the gap between safety and optimality,
	without compromising the safety of the system.\\
	The contributions of this paper are three-fold. First, we formulate the safety problem of RL-based optimal control as the minimization of the desired cost functional subject to a constraint involving the time derivative of the BLF. We
	subsequently formulate the equivalent Lagrangian functional and develop optimal
	control law for the constrained system via the KKT conditions.
	Second, we show that the naive estimation of the Lagrange multiplier fails to
	ensure safety of the underlying system.
	Third, we propose an online estimator for
	the Lagrange multiplier that ensures safety of the RL agent. The proposed
	Actor-Critic-Identifier-Lagrangian algorithm generates a control
action that bridges the gap between safety and performance, i.e., the proposed controller
provides safeguarding action only when needed. When the state trajectories are away
from the unsafe region, the optimizing control action dominates. On the other
hand, when the state approaches the boundary of the safe set, the safeguarding
control action dominates and ensures safety. We argue that
under such an estimation of the Lagrange multiplier, the proposed method is able
to ensure safety while being less conservative than similar methods in the literature.
	To demonstrate the efficacy of the proposed controller, we perform simulation
	studies on two systems with both convex and non-convex state constraints and compare our results
	with similar approaches in the literature.
}


\subsection{Notation}
\label{sec:notation}
We use $\|\cdot\|$ to denote the Euclidean norm for the vectors and the
corresponding induced norm for  matrices. We use $\nabla_{x}$ to denote the
gradient operator with respect to the state $x$, or in other words, $\nabla_x (\cdot) \triangleq \frac{\partial (\cdot)}{\partial x}$. Let
$\mathbb{I}_{n}$ denote the $n \times n$ identity matrix. Let
$\mathcal{L}_{\infty}$ denote the set of all bounded signals.
For a compact set $\mathcal{C}$, we define the sets \(\partial \mathcal{C}\) and
\(\text{Int}(\mathcal{C})\) to be the boundary and the interior of the set
\(\mathcal{C}\), respectively.
We denote the set
of all natural numbers upto and including $n$ by $\mathbb{N}_{n}$. We define
\(\overline{\|(\cdot)\|} \triangleq \sup_{x \in \text{Int}(\mathcal{C})} \|(\cdot)\|\)
for a continuous mapping $(\cdot) : \mathcal{C} \rightarrow \mathbb{R}^{q}$.
$\lambda_{\min}(A)$ denotes the minimum
eigenvalue of the square matrix $A$.
\section{Problem Formulation}
\label{sec:problemFormulation}
In this paper, we consider the state-constrained optimal control problem for a
class of nonlinear systems
\begin{equation}
	\label{eq:plantDynamics}
	\dot{x} = f(x,\theta) + g(x)u, \;\;\; x(t_{0}) = x_{0},
\end{equation}
where \(x(t) \in \mathbb{R}^{n}\) and \(u(t) \in \mathbb{R}^{m}\) are the state
and control vectors, respectively;
\(f: \mathbb{R}^{n} \times \mathbb{R}^{p} \rightarrow \mathbb{R}^{n}\) and
\(g: \mathbb{R}^{n} \rightarrow \mathbb{R}^{n\times m}\) are Lipschitz
continuous functions and \(\theta \in \Theta \subset \mathbb{R}^{p}\) is
a
\textcolor{hcolor}{constant}
unknown parameter vector.
We consider the drift dynamics
\(f(x,\theta)\) to be linear in the parameter \(\theta\), i.e.
\begin{equation}
	\label{eq:driftDynamics}
	f(x,\theta) = Y(x)\theta,
\end{equation}
where we denote the continuous function
\(Y: \mathbb{R}^{n} \rightarrow \mathbb{R}^{n \times p}\) as the
regressor, with $Y(0)=0$. We assume complete knowledge of the functions \(Y(x)\) and \(g(x)\), and
state \(x\) is considered to be measurable. The objective of the safe RL agent is to choose a
control policy \(u(t)\;\;\forall\; t \in [t_{0},\infty)\), that minimizes the cost
functional
\begin{equation}
	\label{eq:costFunctional}
	J(x,t_{0}) = \int_{t_{0}}^{\infty} r(x,u) dt,
\end{equation}
where the positive semi-definite function \(r:\mathbb{R}^{n} \times \mathbb{R}^{m} \rightarrow \mathbb{R}\)
denotes the instantaneous cost function.
In the present work, we solve the optimal regulation problem for the system in
\eqref{eq:plantDynamics} using the cost function defined as
\begin{equation}
	\label{eq:instantaneousCost}
	r(x,u) \triangleq Q(x)+ \frac{1}{2} u^{T} R u,
\end{equation}
where \(Q: \mathbb{R}^{n} \rightarrow \mathbb{R}\) is a positive semi-definite
function of the state with \(Q(0) = 0\) and \(R \in \mathbb{R}^{m\times m}\) is
a positive-definite control effort weight matrix.
In addition to minimizing the cost functional in
\eqref{eq:costFunctional}, the control policy must
also ensure the safety of the system
by ensuring the forward invariance \cite{blanchini1999Automatica} of a user-defined compact set
\(\mathcal{C} \subset \mathbb{R}^{n}\) containing the
origin, i.e.,
\begin{equation}
	\label{eq:originalConstraint}
	x(t_{0}) \in \mathcal{C} \implies x(t) \in \mathcal{C} \;\;\forall\; t \in [t_{0},\infty).
\end{equation}
To ensure satisfaction of \eqref{eq:originalConstraint}, we transform
the set membership constraint to an equivalent formulation using a barrier
Lyapunov function (BLF), as discussed in the subsequent section.
\subsection{Constraint reformulation using barrier Lyapunov functions}
We define a barrier Lyapunov function over the
constraint set \(\mathcal{C}\) as follows
\begin{dfn}[Barrier Lyapunov function \cite{tee2009Automatica}]
	\label{defn:blf}
	A positive-definite continuously
	differentiable function $B_{f}(x): \mathcal{C} \rightarrow \mathbb{R}$
	satisfying
	$B_{f}(0) = 0$,
	$B_{f}(x) > 0 \;\;\forall\; x \in \mathcal{C}\backslash \{0\}$, and
	$\lim_{x\rightarrow \partial \mathcal{C}}B_{f}(x) = \infty$
	is called a barrier Lyapunov function (BLF), if its time derivative along the
	system trajectories is negative semi-definite, i.e., $\dot{B}_f(x) \le 0
		\;\;\forall\; x \in \text{Int}(\mathcal{C})$.
\end{dfn}
We now use the definition of BLF to state the following Lemma which relates the
concepts of BLF and forward invariance of \(\mathcal{C}\).
\begin{lma}
	\label{lem:blf}
	The existence of a BLF $B_{f}$ over the compact constraint set
	$\mathcal{C}$ for a system $\dot{x} = f(x,\theta) +g(x)u$, implies that $\mathcal{C}$
	is forward invariant \cite[Lemma 1]{tee2009Automatica}.
\end{lma}
From Lemma \ref{lem:blf}, we can infer that provided a user-defined BLF \(B_{f}\)
and a control policy \(u(t)\) satisfying \(\dot{B}_{f}|_{u(t)} \le 0\), we can ensure the satisfaction of the constraint \eqref{eq:originalConstraint}.
We can thus reformulate the state constraint
\eqref{eq:originalConstraint} to an equivalent formulation using the BLF \(B_{f}\) as
\begin{equation}
	\begin{aligned}
		\dot{B}_f(x(t))     & \le 0                                                           \\
		\implies
		\nabla_x B_f(x)^{T} & [f(x,\theta) + g(x)u] \le 0 \;\;\forall\; t \in [t_{0},\infty),
	\end{aligned}
\end{equation}
where the time dependence of the signals have been suppressed for notational brevity.
We would utilize this constraint reformulation to obtain a
closed-form solution for the original constrained optimal control problem.
Specifically, we construct a user-defined candidate BLF \(B_{f}(x)\) in such a way that the following
assumption holds.
\begin{assm}
	\label{assum:gradBfConstruction}
	$\exists \; \gamma \in \mathbb{R}_{>0}$
	satisfying
	$\gamma \|\nabla_x B_f(x)\| \ge B_{f}(x)\; \;\;\forall\; x \in \text{Int}(\mathcal{C})$.
	Additionally, $\|\nabla_x B_f(0)\| = 0$.
\end{assm}
The knowledge of the constant \(\gamma\) is used in Section \ref{sec:safety} for obtaining
the largest invariant subset of \(\mathcal{C}\). We provide a few examples of
commonly used BLFs in Table \ref{tab:gamma} that satisfy Assumption \ref{assum:gradBfConstruction}.
\begin{table}[htbp]
	\caption{Value of \(\gamma\) for commonly used BLFs satisfying Assumption \ref{assum:gradBfConstruction}}
	\label{tab:gamma}
	\resizebox{\columnwidth}{!}{
		\begin{tabular}{|c|c|c|}
			\hline
			Safe set (\(\mathcal{C}\))                      & Candidate BLF (\(B_{f}(x)\))                            & \(\gamma\)          \\
			\hline
			\(\{x \in \mathbb{R}^{n}: \|x\| \le \beta  \}\) & \(\ln(\frac{\beta^{2}}{\beta^{2} - x^{T}x})\)           & \(0.5\beta\)        \\
			$\{x \in \mathbb{R}^{n} : |x_{i}| \le a_{i} \;\forall\; i \in \mathbb{N}_{n}  \}$
			                                                & $\sum_{i} \ln(\frac{a_{i}^{2}}{a_{i}^{2} - x_{i}^{2}})$ & $0.5 \max_i(a_{i})$ \\
			\hline
		\end{tabular}}
\end{table}

We now state the main problem formulation for the present work, the BLF-based
state-constrained optimal control problem (BLF-COCP)
\begin{prb}[BLF-COCP]
	\label{prob:equivalent}
	\begin{subequations}
		\begin{align}
			\min_{u(\tau)\;\;\forall\; \tau \in [t_{0},\infty)} \hspace{2pt}
			 & \int_{t_{0}}^{\infty} Q(x(\tau)) + \frac{1}{2} u(\tau)^{T}Ru(\tau) d\tau, \\
			\text{s.t.} \hspace{20pt}
			 & \nabla_x B_f^{T}[f(x,\theta) + g(x) u(t)]\le 0,
			\label{eq:barrierConstraint}                                                 \\
			 & \dot{x} = f(x(t),\theta)+g(x(t))u(t)\;\;\forall\; t \in [t_{0},\infty),   \\
			 & x(t_{0}) \in \mathcal{C}.
			\label{eq:modInitCond}
		\end{align}
	\end{subequations}
\end{prb}
\subsection{Feasibility of BLF-COCP}
We will now show that there exists at least one feasible control policy that
satisfies the constraints of Problem \ref{prob:equivalent}.
We make the
following assumptions on the control effectiveness matrix \(g(x)\) and the
candidate BLF \(B_{f}(x)\)
\begin{assm}
	\label{assum:gFullRank}
	The control effectiveness matrix $g(x)$ satisfies $\lambda_{\min}\big(R_{g}(x)\big) > l_{g} \; \text{and} \; \|R_{g}(x)\| \le \overline{R}_g \;\;\forall\; x \in \mathcal{C}$,
	where $R_{g}(x): \mathbb{R}^{n} \rightarrow \mathbb{R}^{n\times n}$ is defined
	as $ R_{g}(x) \triangleq g(x)R^{-1}g^{T}(x)$ and
	$\overline{R}_g,l_{g} \in \mathbb{R}_{>0}$ are computable positive constants.
\end{assm}
\begin{assm}
	\label{assum:controllability}
	The control matrix $g(x)$ and the barrier Lyapunov function $B_{f}(x)$ satisfy the condition
	$\nabla_x B_f^{T}(x)g(x) \ne 0 \;\;\forall\; x \in \text{Int}(\mathcal{C}) \backslash \{0\}$.
\end{assm}

\begin{rmark}
	\label{rem:regardingControllability}
	Assumption \ref{assum:controllability} may be considered as a controllability condition, extended to incorporate a sense of safety. The assumption may be verified a-priori by appropriately constructing the candidate BLF $B_f(\cdot)$. Additionally, Assumption \ref{assum:controllability} may be relaxed by placing a condition on the unforced drift dynamics $f(x)$ \cite[Remark 2]{cohen2023Automatica}.
\end{rmark}
We now state the following Lemma.
\begin{lma}
	\label{lem:existence}
	Under Assumption \ref{assum:controllability}, there exists a control policy
	$u_{safe}(x) = -\frac{g(x)^{T} \nabla_x B_f(x)}{\|g(x)^{T} \nabla_x B_f(x)\|^{2}}
		\nabla_x B_f^{T}(x)f(x,\theta),$
	satisfying the constraints of Problem \ref{prob:equivalent}\footnote{
		Lemma \ref{lem:existence} can be verified by substituting
		$u_{safe}$ in the LHS of constraint \eqref{eq:barrierConstraint}.
	}.
\end{lma}
From Lemma \ref{lem:existence}, we can conclude that there exists atleast one
feasible control policy satisfying the constraints of Problem
\ref{prob:equivalent}.

\section{Lagrangian-based constrained optimal control}
\label{sec:adp}
Typically, constrained optimization problems are solved by defining a Lagrangian that
transforms the problem into
an equivalent unconstrained optimization problem \cite{boyd2004convex}. In the
same vein, constrained optimal control problems can be solved by adjoining the
performance objective functional with the constraints weighted by a Lagrange multiplier
\cite[Section 3.2]{lewis2012optimal}.
The Hamiltonian
$H: \mathbb{R}^{n} \times \mathbb{R}^{m} \times \mathbb{R}^{n} \to \mathbb{R}$, for the optimal control problem is defined as
\begin{equation}
	H(x,u,\nabla_{x}V_{u}) = r(x,u) + \nabla_{x}V_{u}^{T}(f(x) + g(x)u),
\end{equation}
where $V_{u}: \mathbb{R}^{n} \to \mathbb{R}$ is the value function for the
unconstrained optimal control problem.

The state-constrained optimal control problem (Problem
\ref{prob:equivalent}) is reformulated as a constrained Hamiltonian minimization problem as follows
\begin{prb}
	\label{prob:reformulation}
	\begin{subequations}
		\begin{align}
			u^{*}(x) & = \min_{u} H(x,u,\nabla_{x}V^{*}(x)), \\ \text{s.t.} \hspace{0.7cm} & \nabla_x B_f^{T}[f(x,\theta) + g(x) u(t)]\le 0,
		\end{align}
	\end{subequations}
\end{prb}
where $V^{*}: \mathbb{R}^{n} \to \mathbb{R}$ is the optimal value function for
the constrained Problem \ref{prob:equivalent}. It can be shown that the Problem
\ref{prob:reformulation} and Problem \ref{prob:equivalent} are point-wise
equivalent. To solve Problem \ref{prob:reformulation} and consequently Problem
\ref{prob:equivalent}, we define the Lagrangian (cf. \cite{almubarak2021CDC}) as
\begin{equation*}
	L(x,u,\lambda) \triangleq H(x,u,\nabla_{x}V^{*}(x)) + \lambda \nabla_x B_f^{T}[f(x,\theta) + g(x) u],
\end{equation*}
where the dual variable $\lambda(x) : \mathbb{R}^{n} \to \mathbb{R}$ is the state-dependent Lagrange multiplier.
According to the Lagrangian theory, the solution to the constrained
problem (Problem \ref{prob:reformulation}) is achieved by unconstrained minimization
of $L(\cdot)$ subject to the control law $u$ and the dual variable $\lambda$
\cite{boyd2004convex}.
\textcolor{hcolor}{
	We observe that
	the optimization problem in Problem \ref{prob:reformulation} is
	convex in the decision variable \(u(t)\). Thus,
	the optimal solution
	can be obtained by invoking the
	Karush-Kuhn-Tucker (KKT) conditions \cite[Section 5.3.3]{boyd2004convex} which are both necessary and sufficient
	for optimality. We thus write
}
\begin{subequations}
	\begin{equation}
		\label{eq:firstOrderCondition}
		\frac{\partial}{\partial u}
		\Big[r(x,u)+ \lambda^{*} \nabla_x B_f^{T}(f+gu) +
		\nabla_x V^{*T}(f+gu)\Big]= 0,
	\end{equation}
	\begin{equation}
		\label{eq:complementarySlackness}
		\lambda^{*}(x(t)) \nabla_x B_f^{T}(f + g u^{*})= 0,
	\end{equation}
	\begin{equation}
		\label{eq:dualConstraints}
		\lambda^{*}(x(t))\ge 0 \;\forall\; t \in \mathbb{R}_{\ge t_{0}}.
	\end{equation}
\end{subequations}
Using the first order condition in \eqref{eq:firstOrderCondition} and the
instantaneous cost function from \eqref{eq:instantaneousCost}, we can write the
optimal control law for Problem \ref{prob:equivalent} as
\begin{equation}
	\label{eq:optimalControlLaw}
	\begin{aligned}
		u^{*}(x) & = -R^{-1}g^{T}(x)[\nabla_x V^{*}(x) + \lambda^{*}(x) \nabla_x B_f(x)] \\
		         & = u_{v}^{*}(x) + u_{\lambda}^{*}(x),
	\end{aligned}
\end{equation}
where, the control law $u^{*}(x)$ consists of two components namely the value
function component
$u_{v}^{*}(x) \triangleq -R^{-1}g^{T}(x)\nabla_{x}V^{*}(x)$ and the
safeguarding controller
$u_{\lambda}^{*}(x) \triangleq - \lambda^{*}(x)R^{-1}g^{T}(x)\nabla_{x}B_{f}(x)$.
\textcolor{hcolor}{
	Additionally, we observe that the control law in \eqref{eq:optimalControlLaw}
	satisfies the second-order sufficient conditions as
	$
		\frac{\partial^{2} L}{\partial u^{2}} = R \succ 0.
	$
}
Substituting \eqref{eq:optimalControlLaw} in the complementary slackness
condition \eqref{eq:complementarySlackness} and using the condition on the dual
constraint \eqref{eq:dualConstraints} we obtain the optimal Lagrange multiplier
(cf. \cite{almubarak2021CDC}) as
\begin{equation}
	\label{eq:lambdaStar}
	\lambda^{*}(x) = \begin{cases}
		\max\Big(\frac{C_{s}^{*}(x) }{R_{bf}(x)}, 0\Big), & \text{if } \nabla_x B_f(x) \ne 0, \\
		0,                                                & \text{otherwise},                 \\
	\end{cases}
\end{equation}
where $R_{bf}(x) \triangleq \nabla_x B_f(x)^{T}R_{g}(x)\nabla_x B_f(x)$,
\(R_{g}(x)\) was defined in Assumption \ref{assum:gFullRank} and
\begin{equation}
	C_{s}^{*}(x) \triangleq \nabla_x B_f(x)^{T} f(x,\theta) - \nabla_x B_f(x)^{T} R_{g}(x) \nabla_x V^{*}.\label{eq:CsStar}
\end{equation}
Substituting the values of $u^{*}(x)$ and $\lambda^{*}(x)$ in the definition of the
Lagrangian, we obtain the optimality condition
\begin{equation}
	\begin{aligned}
		\label{eq:optimalityCondition}
		0= \min_{\substack{u(t),\lambda(x)}} & 
		\Big[r(x,u)+ \lambda \nabla_x B_f^{T}[f(x,\theta)+g(x)u]                                                            \\
		                                     & +\nabla_x V^{*T}[f(x,\theta)+g(x)u]\Big] \;\;\forall\; x \in \mathbb{R}^{n}.
	\end{aligned}
\end{equation}
\textcolor{hcolor}{
	\begin{lma}
		\label{lem:optimalControlSafety}
		For the constrained optimal control problem defined in Problem
		\ref{prob:equivalent}, the control law in \eqref{eq:optimalControlLaw}
		with the Lagrange multiplier defined in \eqref{eq:lambdaStar} satisfies
		the following properties.
		\begin{enumerate}[\bf i.]
			\item \textbf{Optimality:} $u^{*}(x)$ is the optimal solution to Problem \ref{prob:equivalent}.
			\item \textbf{Safety:} The set $\mathcal{C}$ is forward invariant for the system
			      in \eqref{eq:plantDynamics}.
			\item \textbf{Stability:} The origin is locally asymptotically stable with
			      respect to the
			      system in \eqref{eq:plantDynamics}.
		\end{enumerate}
	\end{lma}
	\begin{pf*}{Proof}
		\begin{enumerate}[\bf i.]
			\item 
			      Using \eqref{eq:optimalityCondition} and the complementary
			      slackness relation \eqref{eq:complementarySlackness}, we can write
			      \begin{equation*}
				      r(x,u^{*}(x)) + \nabla_{x}V^{*T}[f(x,\theta) + g(x)u^{*}] = 0.
			      \end{equation*}
			      Using the definition of Hamiltonian, we can rewrite the above
			      expression as $H(x,u^{*},\nabla_{x}V^{*}) = 0$, which is the
			      optimality condition (Hamilton-Jacobi-Bellman equation) for infinite horizon optimal
			      control problems
			      \cite{bhasin2013Automatica,almubarak2021CDC}. Thus $u^{*}(x)$
			      is the optimal solution to Problem \ref{prob:equivalent}. \\
			\item 
			      Consider the positive-definite candidate Lyapunov function $B_{f}(x)$. The time derivative of the same along the system trajectories is given by
			      $\dot{B}_{f}(x) = \nabla_x B_f(x)^{T}[f(x,\theta) + g(x)u^{*}(x)]$.
			      Substituting $u^{*}(x)$ and $\lambda^{*}(x)$ from \eqref{eq:optimalControlLaw} and
			      \eqref{eq:lambdaStar} respectively, we write
			      \begin{equation*}
				      \begin{aligned}
					      \dot{B}_{f}(x) = \begin{cases}
						                       C_{s}^{*}(x), & \text{if } C_{s}^{*}(x) \le 0 \;\text{and}\; \nabla_x B_f(x) \ne 0,                                 \\ 0,            & \text{otherwise}.                        \\ \end{cases}
				      \end{aligned}
			      \end{equation*}
			      We observe that the time derivative of the BLF $B_{f}(x)$ can be bound as
			      $
				      \dot{B}_{f}(x) \le 0.
			      $
			      We thus conclude that $B_{f}(x)$ is a valid Barrier Lyapunov function.
			      Invoking Lemma \ref{lem:blf}, we can show that the set $\mathcal{C}$ is forward
			      invariant for the system in \eqref{eq:plantDynamics}. \\
			\item 
			      Consider the candidate Lyapunov function $V^{*}(x)$. Taking
			      the time derivative along the system trajectories we have
			      $\dot{V}^{*}(x) = \nabla_{x}V^{*}(x)^{T}[f(x,\theta) + g(x)u^{*}]$.
			      Using \eqref{eq:complementarySlackness} and \eqref{eq:optimalityCondition} we can write
			      $
				      \dot{V}^{*}(x) = -r(x,u^{*}(x)) < 0 \;\;\forall \; x  \in \text{Int}(\mathcal{C}) \backslash \{0\}.
			      $
			      We observe that $\dot{V}^{*}(\cdot)$ is negative definite
			      in the domain of $\text{Int}(\mathcal{C})$. As a result, the
			      origin is locally asymptotically stable for the system in \eqref{eq:plantDynamics}  \cite{khalil2002nonlinear}.
			      \qed
		\end{enumerate}
	\end{pf*}
}

\begin{rmark}
	\label{rem:switchingOfLambdaStar}
	The term $C_{s}^{*}(x)$ is the time-derivative of the BLF $B_{f}(x)$ under
	the control law $u_{v}^{*}(x)$ (i.e., without the influence of the
	safeguarding controller $u_{\lambda}^{*}(x)$).
	If
	$C_{s}^{*}(x)$ is negative at a particular state $x$, then the value function component of the control
	law ($u_{v}^{*}(\cdot)$) alone suffices to ensure system safety at $x$ and thus,
	the optimal control law switches off the safeguarding controller (i.e.,
	$u_{\lambda}^{*}(x) = 0$). On the other hand, when $C_{s}^{*}(x)$ is positive,
	the safeguarding controller ($u_{\lambda}^{*}(x)$) is switched on, and the
	optimal control law in \eqref{eq:optimalControlLaw} ensures $\dot{B}_{f}(x) \le 0$ at
	state $x$.
\end{rmark}

Since \(V^{*}(x)\) and \(\theta\) are unknown, the terms \(C_{s}^{*}(\cdot)\) and consequently
\(\lambda^{*}(\cdot)\) are unknown. In the subsequent section, we propose an online estimation technique to
approximate these terms, leading to the design of the approximate optimal controller.

\section{Online estimation of the constrained optimal controller}
\label{sec:acil}
Since the value function \(V^{*}(x)\) is unknown and difficult to compute
analytically, we use a single-layer neural network defined over a compact set
\(\mathcal{D} \subset \mathbb{R}^{n}\) containing the origin such that \(\mathcal{C} \subseteq \mathcal{D}\), to
approximate \(V^{*}(x)\). To that end, we
consider a continuously differentiable user-defined basis function
\(\phi(x): \mathcal{D} \rightarrow \mathbb{R}^{b}\) with \(\phi(0) = 0\) and \(\nabla_{x} \phi(0) = 0\). We
parameterize the value function \(V^{*}(x)\) as
\begin{equation}
	\label{eq:LstarDefinition}
	\begin{aligned}
		V^{*}(x) = W^{T}\phi(x) + \epsilon(x),
	\end{aligned}
\end{equation}
where \(W \in \mathbb{R}^{b}\) is the unknown weight
vector, and \(\epsilon: \mathcal{D} \rightarrow \mathbb{R}\) is the function reconstruction error
for the proposed neural network.
\textcolor{hcolor}{
\begin{assm}
  \label{assm:knownUpperBoundW}
An upper-bound of the norm of the neural network weight
$\|W\| \leq \overline{W}$ is known, where $\overline{W} \in \mathbb{R}_{>0}$ is a
positive constant.
\end{assm}
Assumption \ref{assm:knownUpperBoundW} is standard in the literature of
approximate dynamic programming \cite{vamvoudakis2010Automatica,bhasin2013Automatica,kamalapurkar2016Automatica}. The knowledge of
this upper-bound would be subsequently used to establish bounds on the estimates
of $W$.
}

\begin{lma}
	The neural network function reconstruction error $\epsilon(x)$
	and its derivative w.r.t. state are bounded as $\|\epsilon(x)\|\le
		\overline{\epsilon}$ and $\|\nabla_{x}\epsilon(x)\|\le \overline{\epsilon}_{d} \;\;\forall\; x \in \mathcal{D}$
	, where $\overline{\epsilon}, \overline{\epsilon}_{d} \in \mathbb{R}_{>0}$ are positive constants.
	Additionally these bounds can be made arbitrarily close to zero by choosing
	an appropriate basis function $\phi$ and increasing the number of neurons \cite{kreinovich1991NN}.
\end{lma}
Further, we consider the drift dynamics in \eqref{eq:driftDynamics} to be
parameterized by an unknown parameter \(\theta\), which, we estimate by
\(\hat{\theta}(t) \in \mathbb{R}^{p}\). Thus, the estimated drift dynamics is defined as
\begin{equation}
	\label{eq:driftDynamicsEstimate}
	\hat{f}(x,\hat{\theta}) = Y(x) \hat{\theta}.
\end{equation}
We design two estimates of the unknown weight vector $W$ in \eqref{eq:LstarDefinition}, i.e.,
$\hat{W}_a(t) \in \mathbb{R}^{b}$ that is used to estimate the control action, and
$\hat{W}_c(t) \in \mathbb{R}^{b}$ that is used to estimate the value function.
Thus, we write the estimate for the value function from
\eqref{eq:LstarDefinition} and control law from \eqref{eq:optimalControlLaw} as
\begin{equation}
	\label{eq:LHat}
	\hat{V}(x,\hat{W}_c) = \hat{W}_c^{T}\phi(x),
\end{equation}
\begin{equation}
	\label{eq:finalControlLaw}
	\begin{aligned}
		\hat{u}(x,\hat{W}_a,\hat{\theta}) = & 
		-R^{-1}g^{T}(x)[
		\nabla_{x} \phi(x)^{T}\hat{W}_a        \\&
				+ \hat{\lambda}(x,\hat{W}_{a},\hat{\theta}) \nabla_x B_f(x)],
	\end{aligned}
\end{equation}
respectively, where
\(\hat{\lambda}(x,\hat{W}_{a},\hat{\theta}): \mathbb{R}^{n} \times \mathbb{R}^{b} \times \mathbb{R}^{p} \rightarrow \mathbb{R}\)
is a subsequently designed online estimate of the
Lagrange multiplier defined in \eqref{eq:lambdaStar}.

We now propose an online on-policy\footnote{
	A reinforcement learning algorithm is said to be on-policy, if the data used to train the policy is generated by applying the same policy\cite{sutton2018reinforcement}.
} RL algorithm consisting of the following components.
\begin{itemize}
	\item \textbf{Actor} - The update law for \(\hat{W}_a\) for estimating the
	      control law, based on the minimization of the actor parameter estimation error
	      $\tilde{W}_a(t) \triangleq W - \hat{W}_a(t)$.
	\item \textbf{Critic} - The update law for \(\hat{W}_c\) for estimating the
	      value function, based on the minimization of the critic parameter estimation error
	      $\tilde{W}_c(t) \triangleq W - \hat{W}_c(t)$.
	\item \textbf{Identifier} - The update law for \(\hat{\theta}\) for estimating
	      the system drift dynamics, based on the minimization of the parameter estimation
	      error $\tilde{\theta}(t) \triangleq \theta - \hat{\theta}(t)$.
	\item \textbf{Lagrangian} - The algebraic expression for \(\hat{\lambda}\)
	      estimating the optimal Lagrange multiplier, based on the minimization of the error
	      $\tilde{\lambda}(x) \triangleq \lambda^{*}(x) - \hat{\lambda}(x)$.
\end{itemize}
We now provide details of each of the components of the proposed
Actor-Critic-Identifier-Lagrangian (ACIL) algorithm in the following subsections.

\subsection{Online estimation of Lagrange multipliers}
We observe that the Lagrange multiplier derived in Section \ref{sec:adp}
involves the $\max(\cdot,\cdot)$ function in its expression which is similar in
spirit to the rectified linear unit (ReLU) function in the machine learning
parlance \cite{nair2010rectified}. Intuitively, as discussed in Remark
\ref{rem:switchingOfLambdaStar}, the Lagrange multiplier switches on/off the
safeguarding control law based on the safety requirements of the current state.
However, such a switching technique relies on perfect knowledge of the optimal
value function $V^{*}(\cdot)$ and the drift dynamics parameter $\theta$. Under
uncertainty in these terms, it may not be straightforward to ensure safety of
the system. To demonstrate this, we construct a Lagrange multiplier estimate
using the certainty-equivalence principle (i.e., replacing unknown terms in
$\lambda^{*}(x)$ with their corresponding estimates) as
\begin{equation}
	\label{eq:lambdaNaive}
	\hat{\lambda}_{n}(x,\hat{W}_{a},\hat{\theta}) = \begin{cases}
		\max\Big(\frac{\hat{C}_{s}(x,\hat{W}_{a},\hat{\theta}) }{R_{bf}(x)}, 0\Big), & \text{if } \nabla_x B_f(x) \ne 0, \\
		0,                                                                           & \text{otherwise},                 \\
	\end{cases}
\end{equation}
where
\begin{equation}
	\begin{aligned}
		\hat{C}_{s}(x, \hat{W}_a,\hat{\theta}) & \triangleq
		\nabla_x B_f(x)^{T}f(x,\hat{\theta})                \\&
		- \nabla_x B_f(x)^{T}R_{g}(x) \nabla_{x} \phi(x)^{T} \hat{W}_a,
	\end{aligned}
\end{equation}
is the estimate of $C_{s}^{*}(\cdot)$. We consider the
candidate Lyapunov function $B_{f}(x)$ and compute the time derivative
along the system trajectories under control \eqref{eq:finalControlLaw} with
the estimate $\hat{\lambda}_{n}$ from \eqref{eq:lambdaNaive}, to write
\begin{equation}
	\label{eq:pathologicalCase}
	\dot{B}_{f} = C_{s}^{*} + \nabla_{x}B_{f}^{T}R_{g}(\nabla_{x}\epsilon + \nabla_{x}\phi^{T}\tilde{W}_{a})
	- \max(\hat{C}_{s},0).
\end{equation}
During the transient phase of RL training, when the estimate $\hat{C}_{s}(\cdot)$ may
not be close to $C_{s}^{*}(\cdot)$, there may exist some
scenarios where $C_{s}^{*} > 0$ while $\hat{C}_{s} < 0$. In such a
case, $\dot{B}_{f} > 0$ and consequently the BLF grows unbounded, and consequently,
the RL agent fails to ensure safety of the system.

It is evident that a better online estimate for the Lagrange multiplier is
required to ensure safety under parametric uncertainties. To that end, we
propose to use a smooth approximation to the ReLU function to provide a degree
of robustness against the uncertainty in the estimation of $C_{s}^{*}(\cdot)$. We construct
a continuously differentiable function
$\sigma: \mathbb{R} \rightarrow \mathbb{R}$ with the following properties-
\begin{properties}
	\item Both $\sigma(\cdot)$ and its derivative $\sigma'(\cdot)$ are monotonically
	increasing.
	\item $\sigma(z) \ge \max(z,0) \;\;\forall\; z \in \mathbb{R}$.
	\item $\sigma(\cdot)$ asymptotically approaches the $\max(\cdot,0)$ function. In
	other words,
	$\lim_{z \rightarrow -\infty} \sigma(z) - \max(z,0) = \lim_{z \rightarrow \infty} \sigma(z) - \max(z,0) = 0$.
\end{properties}
\begin{rmark}
	The properties P1 and P3 ensure
	that $\sigma(\cdot)$ is a smooth convex approximation of ReLU function.
	Additionally, using properties P1 and P3, we can show that $0 \leq \sigma'(x) \leq 1 $
	and $\sigma''(x) \geq 0 \;\;\forall\; x \in \mathbb{R}$.
	Property P2 is useful in providing a negative semi-definite term in the
	subsequent Lyapunov analysis, and thus facilitates safety under parametric uncertainties.
\end{rmark}

Using the properties P1 to P3, we now state an important Lemma which would be
subsequently used in proving safety of the closed loop system.
\begin{lma}
	\label{lem:linearityOfSoftplus}
	The function $f_{\sigma}:(0,\infty) \rightarrow \mathbb{R}$ defined as
	$f_{\sigma}(z) \triangleq \sigma(c/z) z$,
	where $c \in \mathbb{R}_{>0}$ is a constant, can be upper-bound as
	$f_{\sigma}(z) \le \sigma_{0} z + c
		\;\;\forall\; z \in (0,\infty)$, where
	$\sigma_{0} \triangleq \sigma(0)$.\footnote{See Appendix for the proof.}
\end{lma}
Based on this smooth approximation of the ReLU function, we propose an online
estimate \(\hat{\lambda}(\cdot)\) of the Lagrange multiplier \(\lambda^{*}(\cdot)\) from
\eqref{eq:lambdaStar} as
\begin{equation}
	\label{eq:lambdaHat}
	\begin{aligned}
		\hat{\lambda}(x,\hat{W}_a,\hat{\theta}) = \sigma\Big(
		\frac{\hat{C}_{s}(x,\hat{W}_a,\hat{\theta})}{R_{bf}(x)
				+ k_{sb}}\Big)
		+ k_{so}\varsigma,
	\end{aligned}
\end{equation}
where $\varsigma \triangleq \frac{2\sigma_{0}\overline{R}_g}{k_{so}l_g}+1$ is a
computable positive constant, $k_{so}\in \mathbb{R}_{>0}$ and
$k_{sb} \in \mathbb{R}_{>0}$ are user-defined constants. Ideally, the constants $k_{so}$ and $k_{sb}$
must be close to zero to better approximate the structure of $\lambda^{*}(\cdot)$.
We show the safety guarantees of the proposed controller in Section \ref{sec:safety}.
\begin{rmark}
	The introduction of the function $\sigma(\cdot)$ for the estimation of the Lagrange multiplier
	provides robustness against the uncertainty in $C_{s}^{*}$.
	The proposed Lagrange multiplier estimate ensures that even in the pathological
	case considered before ($C_{s}^{*} > 0$ while $\hat{C}_{s} < 0$), the
	safeguarding control term $u_{sg}$ is non-zero and contributes a stabilizing
	term to \eqref{eq:pathologicalCase}, thus guaranteeing safety under parametric uncertainties. We perform a detailed safety
	analysis in Section \ref{sec:safety}.
\end{rmark}

\begin{rmark}
	Compared to \cite{cohen2023Automatica}, where the gain of the safeguarding controller is constant, the
	proposed estimator of the Lagrange multiplier includes a state-varying gain that
	ensures safety under all conditions of the state. Further, in the scenario when the
	value-function based control ($u_{v}$) suffices to ensure safety, we argue that the
	proposed strategy would be less conservative than \cite{cohen2023Automatica}, in terms of reducing the control
	effort due to the safeguarding controller.
	
\end{rmark}
\begin{figure}[htpb]
	\centering
	\includegraphics[width=0.8\linewidth]{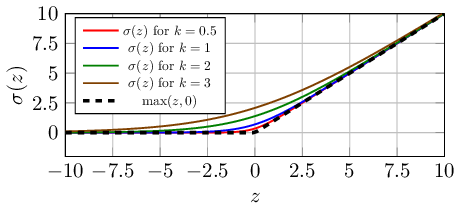}
	\vspace*{-0.3cm}
	\caption{Softplus\cite{nair2010rectified} function}
	\label{fig:softplus}
\end{figure}
One notable example of a function satisfying Properties P1-P3 can be found in the
machine learning literature, namely the ``softplus'' function
\cite{nair2010rectified} defined as
\begin{equation}
	\label{eq:softplus}
	\begin{aligned}
		\sigma(z) \triangleq k \ln(1+ \exp(z/k)),
	\end{aligned}
\end{equation}
where $k \in \mathbb{R}_{>0}$ is a user-defined gain.
We provide the plot for the softplus function for various values of $k$ in Fig.
\ref{fig:softplus}.
We observe that as the value of the gain $k$ decreases, the softplus function
becomes a better approximation for the ReLU function.
\subsection{Actor-Critic design based on simulation of experience}
We now extend the Actor-Critic design from \cite{kamalapurkar2016Automatica}
involving a ``simulation of experience'' paradigm to the
training of online estimators for the value function of constrained optimal
control problem.
The objective of the critic component of the algorithm is to minimize the
Bellman Error (BE)
\(\delta: \mathbb{R}^{n} \times \mathbb{R}^{b}\times \mathbb{R}^{b} \times \mathbb{R}^{p} \rightarrow \mathbb{R}\)
defined as
\begin{equation}
	\label{eq:BE}
	\begin{aligned}
		& \delta(x,\hat{W}_c,\hat{W}_a,\hat{\theta}) \triangleq r(x,u(x,\hat{W}_a,\hat{\theta})) +\big(\nabla_x \hat{V}(x,\hat{W}_c)                               \\ & +\hat{\lambda}(x,\hat{W}_a,\hat{\theta}) \nabla_x B_f(x) \big)^{T} [f(x,\hat{\theta}) + g(x)u(x,\hat{W}_a,\hat{\theta})], \end{aligned}
\end{equation}
which is equivalent to the error in
estimating \eqref{eq:optimalityCondition} by the corresponding estimates for the
ideal parameters. The authors of
\cite{kamalapurkar2016Automatica} observed that the BE can be evaluated
independent of the state trajectory and proposed to evaluate the BE at so called
``Bellman extrapolation points'', to enable a virtual exploration of the state
space. In this paper, we make a slight modification to that framework by
considering the set of functions
\(\{X_{i}:\mathbb{R}^{n} \times [t_{0},\infty) \rightarrow \mathcal{C} \}_{i=1}^{N}\),
such that each \(X_{i}(\cdot)\) maps the current state to a point
\(x_{i}\in \mathcal{C}\).
\textcolor{hcolor}{
	At each time instant, the proposed ACIL algorithm generates a set of ``virtual
	states'' where it evaluates the value of Bellman error.
}
Thus, the critic evaluates the BE at each extrapolation point
\begin{equation}
	\label{eq:BEi}
	\begin{aligned}
		 & \delta_{i}(x_{i},\hat{W}_c,\hat{W}_a,\hat{\theta}) \triangleq r(x_{i},u(x_{i},\hat{W}_a,\hat{\theta})) +\big(\nabla_x \hat{V}(x_{i},\hat{W}_c) \\ & +\hat{\lambda}(x_{i},\hat{W}_a,\hat{\theta}) \nabla_x B_f(x_{i}) \big)^{T}
			[f(x_{i},\hat{\theta}) + g(x_{i})u(x_{i},\hat{W}_a,\hat{\theta})],
	\end{aligned}
\end{equation}
where \(x_{i}(t) = X_{i}(x(t),t) \;\;\forall\; i \in \mathbb{N}_{N}\).
\textcolor{hcolor}{
	\begin{rmark}
		A simple example of the set of functions $X_i(\cdot)$
		$$X_i(x,t) =
			\begin{cases}
				x + r_{bi}, & \text{if} \;\; x+ r_{bi} \in \text{Int}(\mathcal{C}), \\
				r_{s},      & \text{otherwise},\end{cases}
		$$
		where $r_{bi} \in \mathbb{R}^{n}$ is a perturbation vector and $r_{s} \in \text{Int}(\mathcal{C})$ is a fallback point.
		It is worth noting that the Bellman extrapolation points are trajectory dependent.
	\end{rmark}
}
The critic uses the information of the gradient of the Bellman Errors w.r.t.
\(\hat{W}_c\) to update its estimate of the neural network parameter. Specifically,
the critic computes the signals
\(\omega(t),\omega_{i}(t) \in \mathbb{R}^{b} \;\forall\; i \in \mathbb{N}_{N}\) defined as
$\omega(t) \triangleq \frac{\partial }{\partial \hat{W}_c}
	\delta(\cdot)$ and
$\omega_{i}(t) \triangleq \frac{\partial}{\partial \hat{W}_c}
	\delta_{i}(\cdot)$ respectively.
The critic subsequently uses the BEs computed in \eqref{eq:BE} and
\eqref{eq:BEi} to improve the estimate \(\hat{W}_c(t)\) using a recursive least
squares-based update law (RLS update law) as
\vspace{-0.4cm}
\begin{equation}
	\label{eq:critic}
	\begin{aligned}
		\dot{\hat{W}}_c(t) = -\eta_{c1} \Gamma(t) \frac{\omega(t)}{\rho(t)^{2}} \delta(\cdot)
		-\frac{\eta_{c2}}{N} \Gamma(t) \sum_{i=1}^{N}
		\frac{\omega_{i}(t)}{\rho_{i}(t)^{2}} \delta_{i}(\cdot),
	\end{aligned}
\end{equation}
where
$\rho(t) \triangleq \sqrt{1 + \nu \omega(t)^{T}\Gamma(t)\omega(t)}$ and
$\rho_{i}(t) \triangleq \sqrt{1 + \nu \omega_{i}(t)^{T}\Gamma(t)\omega_{i}(t)}$
are normalizing factors with
\(\nu, \eta_{c1},\eta_{c2} \in \mathbb{R}_{>0}\) being user-defined positive gains.
The least squares gain matrix \(\Gamma(t) \in \mathbb{R}^{b \times b}\) in
\eqref{eq:critic} is updated according to the update law
\begin{equation}
	\label{eq:gammaUpdate}
	\begin{aligned}
		\dot{\Gamma}(t) & = \beta \Gamma(t) - \eta_{c1} \Gamma(t) \frac{\omega(t)\omega(t)^{T}}{\rho^{2}(t)} \Gamma(t)                                                                                      \\ & - \frac{\eta_{c2}}{N} \Gamma(t) \sum_{i=1}^{N} \frac{\omega_{i}(t)\omega_{i}(t)^{T}}{\rho_{i}^{2}(t)} \Gamma(t), \; \Gamma(t_0) = \Gamma_{0}, \end{aligned}
\end{equation}
where \(\beta \in \mathbb{R}_{>0}\) is a constant forgetting factor and \(\Gamma_{0} \in
\mathbb{R}^{b \times b}\) is a positive-definite initial gain matrix.

\begin{assm}[Excitation conditions \cite{kamalapurkar2016Automatica}]
	\label{assum:excitationConditions}
	There exist constants $T \in \mathbb{R}_{>0}$ and $c_{1},c_{2},c_{3} \in \mathbb{R}_{\ge 0}$ such that
	the conditions
	$c_{1} \mathbb{I}_{b} \le \int_{t_{0}}^{t_{0}+T} \psi(\tau) d \tau $,
	$c_{2} \mathbb{I}_{b} \le
		\inf_{\tau \in \mathbb{R}_{\ge 0}} \frac{1}{N} \sum_{i=1}^{N}
		\psi_{i}(\tau)$,
	$c_{3} \mathbb{I}_{b} \le
		\frac{1}{N}\int_{t_{0}}^{t_{0}+T} \sum_{i=1}^{N}\psi_{i}(\tau) d \tau$
	hold, where $\psi(\tau) \triangleq \frac{\omega(\tau) \omega(\tau)^{T}}{\rho^{2}(\tau)}$,
	$\psi_{i}(\tau) \triangleq \frac{\omega_{i}(\tau) \omega_{i}(\tau)^{T}}{\rho_{i}^{2}(\tau)}$
	and atleast one of the constants $c_{1},c_{2},c_{3}$ is strictly
	positive\footnote{This assumption is the relaxation for the standard persistence
		of excitation (PE) condition in the theory of adaptive control, which is difficult to
		verify online.
		Assumption \ref{assum:excitationConditions} can
		be verified online by increasing the number (N) of extrapolation points considered
		\cite{kamalapurkar2016Automatica}.}.
\end{assm}
Under Assumption \ref{assum:excitationConditions} and provided
\(\lambda_{\min}(\Gamma^{-1}(t_{0})) > 0\), it can be shown that
\eqref{eq:gammaUpdate} ensures that the gain matrix \(\Gamma(t)\) satisfies
$
	\underline{\Gamma} \mathbb{I}_{b}
	\preccurlyeq \Gamma(t) \preccurlyeq
	\overline{\Gamma} \mathbb{I}_{b},$
where \(\preccurlyeq\) denotes the semi-definite ordering of square matrices and
\(\underline{\Gamma}, \overline{\Gamma} \in \mathbb{R}_{>0}\) are positive constants
with \(\overline{\Gamma} > \underline{\Gamma}\) \cite[Lemma 1]{kamalapurkar2016Automatica}.
We now design the update law for the actor parameter \(\hat{W}_a\) as
\begin{equation}
	\label{eq:actor}
	\begin{aligned}
		\dot{\hat{W}}_a(t) & = \textcolor{hcolor}{\proj_{\Omega_{a}}}\Big(\eta_{a1} (\hat{W}_c(t) - \hat{W}_a(t)) -\eta_{a2} \hat{W}_a(t)                                                                                \\ & +\eta_{c1} \frac{R_{s}(t)^{T} \hat{W}_a(t) \omega(t)^{T}}{4 \rho(t)} \hat{W}_c(t) \\ & +\frac{\eta_{c2}}{N} \sum_{i=1}^{N} \frac{R_{si}(t)^{T} \hat{W}_a(t) \omega_{i}(t)^{T}}{4 \rho_{i}(t)} \hat{W}_c(t) \Big), \end{aligned} \end{equation}
where  \(\eta_{a1}, \eta_{a2} \in \mathbb{R}_{>0}\) are
user-defined positive gains
and
\begin{equation}
	\begin{aligned}
		R_{s}(t)
		 & \triangleq \nabla_{x} \phi(x(t)) R_{g}(x(t)) \nabla_{x} \phi(x(t))^{T},             \\
		R_{si}(t)
		 & \triangleq \nabla_{x} \phi(x_{i}(t)) R_{g}(x_{i}(t)) \nabla_{x} \phi(x_{i}(t))^{T}, \\
	\end{aligned}
\end{equation}
where \(R_{g}(x)\) was defined in Assumption \ref{assum:gFullRank}.
\textcolor{hcolor}{The operator \(\proj_{\Omega_{a}}(\cdot)\) in \eqref{eq:actor} denotes the
projection operator \cite{lavretsky2013robust} that keeps the actor estimate
bounded inside
$\Omega_{a} \triangleq \{w \in \mathbb{R}^{b}: \|w\| \le \overline{W}\}$,
i.e., $\|\hat{W}_a(t)\| \le \overline{W} \;\;\forall\; t \in [t_{0},\infty),$
where \(\overline{W} \in \mathbb{R}_{>0}\) is the known upper-bound of the norm of
the neural network weight $W$ as defined in Assumption \ref{assm:knownUpperBoundW}.
Additionally,
we can bound the error in the actor parameter \(\tilde{W}_a(t)\) as
$\|\tilde{W}_a(t)\| \le \overline{W}_{at} \triangleq 2\overline{W} \;\;\forall\; t \in [t_{0},\infty),$
where  \(\overline{W}_{at} \in \mathbb{R}_{>0}\) is a positive
constant.
}
\subsection{Identifier Design}
\label{subsec:identifier}
Since the Lagrange multiplier estimate
$\hat{\lambda}(x,\hat{W}_{a},\hat{\theta})$ uses the estimate of the system
drift dynamics, we use online system identification techniques to learn the unknown
parameter $\theta$ online from trajectory data.
\textcolor{hcolor}{
Specifically, the estimated parameter $\hat{\theta}$ in \eqref{eq:driftDynamicsEstimate} is modified according to
the general class of update laws of the form
\begin{equation}
	\dot{\hat{\theta}}(t) = f_{\theta}(\hat{\theta}(t),t),
	\label{eq:identifierUpdate}
\end{equation}
where
$f_{\theta}(\cdot) : \mathbb{R}^{p} \times \mathbb{R}_{\ge t_{0}} \rightarrow \mathbb{R}^{p}$
denotes a family of continuous functions given by any of the methods in
\cite{kamalapurkar2016Automatica,roy2017TAC,parikh2019integral}. It can be
shown that under the update law in \eqref{eq:identifierUpdate}, there exists a
Lyapunov function
}
$V_{\theta}(\tilde{\theta},t): \mathbb{R}^{p} \times \mathbb{R}_{\ge t_{0}} \rightarrow \mathbb{R}$
such that
$\dot{V}_{\theta}(\tilde{\theta},t) \le -k_{\theta} \|\tilde{\theta}\|^{2}$,
where $k_{\theta} \in \mathbb{R}_{>0}$ is a positive constant. Consequently, the
following bounds can be established
$\|\hat{\theta}(t)\| \le \overline{\theta}_{h}$,
$\|\tilde{\theta}(t)\| \le \overline{\theta}_{t} \triangleq \overline{\theta} + \overline{\theta}_{h}  \;\;\forall\; t \in [t_{0},\infty)$,
where \(\overline{\theta} \triangleq \|\theta\|\), \(\overline{\theta}_{h}\) and
\(\overline{\theta}_{t}\) are positive constants. We perform a combined Lyapunov
analysis of the actor, critic and the identifier components of the algorithm in
Section \ref{sec:lyap}.

\section{Safety Analysis}
\label{sec:safety}
We state the following lemma which we use in the subsequent Lyapunov analysis
\begin{lma}
	\label{lem:beautifulBound}
	Given a function $\xi(x): \mathbb{R}^{n} \rightarrow \mathbb{R}^{n}$ such that $\|\xi(x)\| \le \Xi \;\;\forall\; x \in \text{Int}(\mathcal{C})$, the bound
	$\Big\|\sigma\Big(\frac{\nabla_x B_f^{T}\xi}{R_{bf}(x)}\Big)\nabla_x B_f\Big\|
		\le
		\sigma_{0}\|\nabla_x B_f\| + \frac{\Xi}{l_{g}} \;\;\forall\; x \in \text{Int}(\mathcal{C})\backslash \{0\}$
	can be established,
	where $\sigma_{0}$ and $l_{g}$ were defined in Lemma \ref{lem:linearityOfSoftplus} and Assumption
	\ref{assum:gFullRank}, respectively
	\footnote{Lemma \ref{lem:beautifulBound} can be shown by utilizing the
		monotonicity properties of the $\sigma$ function and the subsequent trivial application of Lemma
		\ref{lem:linearityOfSoftplus}. Thus, the detailed proof is omitted here for space constraints. }.
\end{lma}

\begin{thm}[\textcolor{hcolor}{Safety certification}]
	\label{thm:safety}
	Provided the actor ($\hat{W}_{a}$), the critic
	($\hat{W}_c$), the drift dynamics parameter estimate ($\hat{\theta}$), and
	the Lagrange multiplier estimate ($\hat{\lambda}$) are updated according to the
	laws detailed in Section \ref{sec:acil}; the estimated optimal control law
	$\hat{u}(x,\hat{W}_{a},\hat{\theta})$ from \eqref{eq:finalControlLaw} ensures
	that the set $\mathcal{C}$ is forward invariant for the system
	in \eqref{eq:plantDynamics}.
	Additionally, the control input $\hat{u}$ is bounded.
\end{thm}
\vspace*{-0.5cm}
\begin{pf*}{Proof}
	For ensuring forward invariance of \(\mathcal{C}\) w.r.t. the system dyanamics we consider
	the positive-definite candidate Lyapunov function \(B_{f}(x)\). The time
	derivative along the system trajectory is given by
	\begin{equation}
		\label{eq:timederLyap}
		\dot{B}_{f}(x) = \nabla_x B_f^{T}(x) [f(x,\theta) + g(x) \hat{u}(x,\hat{W}_a,\hat{\theta})].
	\end{equation}
	Adding and subtracting \(\nabla_x B_f^{T}(x) g(x) u^*(x)\) from the right-hand side of
	\eqref{eq:timederLyap} and using Lemma \ref{lem:optimalControlSafety} we have
	\begin{equation}
		\begin{aligned}
			\dot{B}_{f}(x) \le -\nabla_x B_f(x)^{T}  g(x) [u^{*}(x) - \hat{u}(x,\hat{W}_{a}, \hat{\theta})].
		\end{aligned}
	\end{equation}
	Substituting \(u^{*}\) from \eqref{eq:optimalControlLaw} and \(\hat{u}\) from
	\eqref{eq:finalControlLaw} we have
	\begin{equation}
		\label{eq:beforeLambdaS}
		\begin{aligned}
			\dot{B}_{f} & \le
			\nabla_x B_f(x)^{T} R_{g}(x)
			[\nabla_{x} \phi(x)^{T} \tilde{W}_a + \nabla_{x}\epsilon]
			+\tilde{\lambda} R_{bf}(x).
		\end{aligned}
	\end{equation}
	We now define an intermediate Lagrange multiplier estimate
	\(\lambda_{s}: \mathbb{R}^{n}\rightarrow \mathbb{R}_{>0}\) as
	\begin{equation}
		\label{eq:lambdaS}
		\lambda_{s}(x) \triangleq \sigma\Big(\frac{C_{s}^{*}(x)}{R_{bf}(x)}\Big).
	\end{equation}
	We now add and subtract \(\lambda_{s}(x) R_{bf}(x)\) from the right-hand side
	of \eqref{eq:beforeLambdaS} to obtain
	\begin{equation}
		\begin{aligned}
			& \dot{B}_{f}(x) \le \nabla_x B_f(x)^{T} R_{g}(x) (\nabla_{x} \phi(x)^{T} \tilde{W}_a + \nabla_{x}\epsilon)              \\ & + [\lambda^{*}(x) - \lambda_{s}(x) + \lambda_{s}(x) - \hat{\lambda}(x,\hat{W}_{a},\hat{\theta})] R_{bf}(x). \end{aligned}
	\end{equation}
	Using Assumption \ref{assum:gFullRank}, it can be shown that
	$R_{bf}(x) \ge 0 \; \forall \; x \in \text{Int}(\mathcal{C})$. Furthermore,
	using Property P2 we can bound
	\([\lambda^{*}(\cdot) -\lambda_{s}(\cdot)] R_{bf}(x) \le 0\),
	and write
	\begin{equation}
		\label{eq:beforeTaylor}
		\begin{aligned}
			 & \dot{B}_{f}(x) \le \nabla_x B_f(x)^{T} R_{g}(x) (\nabla_{x} \phi(x)^{T} \tilde{W}_a + \nabla_{x}\epsilon) \\ & + [\lambda_{s}(x) - \hat{\lambda}(x,\hat{W}_{a},\hat{\theta})] R_{bf}(x).
		\end{aligned}
	\end{equation}
	Substituting the values of \(\lambda_{s}(\cdot)\) from \eqref{eq:lambdaS} and
	\(\hat{\lambda}(\cdot)\) from \eqref{eq:lambdaHat} we have
	\begin{equation}
		\begin{aligned}
			 & \dot{B}_{f} \le \nabla_x B_f^{T} R_{g} (\nabla_{x} \phi^{T} \tilde{W}_a + \nabla_{x}\epsilon) -k_{so}\varsigma R_{bf}(x) \\ & +\sigma\Big(\frac{C_{s}^{*}(x)}{R_{bf}(x)} \Big) R_{bf}(x) - \sigma\Big(\frac{\hat{C}_{s}(x,\hat{\theta},\hat{W}_{a})}{R_{bf}(x) + k_{sb}} \Big) R_{bf}(x).
		\end{aligned}
	\end{equation}
	Using the fact that $\sigma(\cdot)$ is monotonically increasing, we write
	\begin{equation}
		\Big| \sigma\Big(\frac{\hat{C}_{s}(\cdot)}{R_{bf}(x) + k_{sb}}\Big)\Big| \le
		\sigma\Big(\frac{|\hat{C}_{s}(\cdot)|}{R_{bf}(x)}\Big).
	\end{equation}
	Using Lemma \ref{lem:beautifulBound} and Assumption \ref{assum:gFullRank} we obtain the following bound
	\begin{equation}
		\begin{aligned}
			 & \dot{B}_{f} \le
			\overline{R}_g (\overline{\phi}_{d} \overline{W}_{at} + \overline{\epsilon}_{d})\|\nabla_x B_f\|
			-k_{so}\varsigma l_{g} \|\nabla_x B_f\|^{2} \\
			 & +
			\Big(
			2\sigma_{0}\|\nabla_x B_f\|
			+ \frac{2 \overline{f} + \overline{\phi}_{d} \overline{W}_{at} + \overline{\epsilon}_{d}}{l_{g}}
			\Big)
			\overline{R}_g\|\nabla_x B_f\|.
		\end{aligned}
	\end{equation}
	Substituting \(\varsigma\) from \eqref{eq:lambdaHat} yields
	\begin{equation}
		\begin{aligned}
			\dot{B}_{f} & \le
			\overline{\chi}\|\nabla_x B_f\|
			-k_{so}l_{g} \|\nabla_x B_f\|^{2},
		\end{aligned}
	\end{equation}
	where
	$\overline{\chi} \triangleq
		\frac{\overline{R}_g}{l_{g}}
		[2\overline{f} + (1 + l_{g})(\overline{\phi}_{d} \overline{W}_{at} + \overline{\epsilon}_{d})]$
	and $\overline{f} \triangleq \overline{\|Y \overline{\theta}_{h}\|} $
	are positive constants. Completing the squares and using
	Assumption \ref{assum:gradBfConstruction} we have
	\begin{equation}
		\label{eq:safetyLyapNegative}
		\begin{aligned}
			\dot{B}_{f} & \le
			-\frac{k_{so}l_{g}}{2\gamma^{2}} B_{f}^{2}
			+ \frac{\overline{\chi}^2}{2 k_{so}l_g}.
		\end{aligned}
	\end{equation}
	We observe that the time derivative of \(B_{f}\) is negative outside the compact
	set
	$\Omega_{b} = \{x \in \mathbb{R}^{n}: B_{f}(x) \le \overline{B}_{f}\}$,
	where \(\overline{B}_{f} \triangleq \frac{\gamma \overline{\chi}}{k_{so}l_{g}}\) is a finite positive constant.
	Under \eqref{eq:safetyLyapNegative} the signal \(B_{f}(x(t))\) can be bounded as
	\begin{equation}
		\begin{aligned}
			B_{f}(x(t)) \le \max(B_{f}(x(t_{0})),\overline{B}_{f}), \;\;\forall\; t \in [t_{0},\infty).
		\end{aligned}
	\end{equation}
	Since \(x(t_{0}) \in \mathcal{C}\) implies that \(B_{f}(x(t_{0}))\) is finite, thus the
	signal \(B_{f}(x(t)) \in \mathcal{L}_{\infty} \;\;\forall\; t \in [t_{0},\infty)\). Since the value of
	the BLF along the system trajectories is bounded, then by definition of
	\(B_{f}(x)\), at no point in time the state trajectory intersects the boundary of
	the safe set (namely \(\partial \mathcal{C}\)) \cite[Lemma 1]{tee2009Automatica}. In other
	words the state \(x(t) \in \mathcal{C} \;\;\forall\; t \in [t_{0},\infty)\) and thus the set \(\mathcal{C}\)
	is forward invariant for the system \eqref{eq:plantDynamics} under the control
	law \eqref{eq:finalControlLaw}.
	
	Since \(B_{f}(x)\) is continuously differentiable in state \(x\), the gradient
	\(\nabla_x B_f(x)\) is a continuous function of state in the compact set \(\Omega_{b}\).
	Thus, the norm of the gradient of the barrier function \(\nabla_x B_f(x(t))\) is bounded along the
	system trajectories by
	\begin{equation}
		\|\nabla_x B_f(x(t))\| \le \overline{B}_{d} \;\;\forall\; t \in [t_{0},\infty),
	\end{equation}
	where \(\overline{B}_{d} \in \mathbb{R}_{>0}\) is a positive constant. Using Lemma
	\ref{lem:beautifulBound}, we can obtain the bound
	$\|\hat{\lambda}(\cdot) \nabla_x B_f(x(t))\| \le \overline{\lambda}_{b} \;\;\forall\; t \in \mathbb{R}_{\ge t_{0}}$,
	where $\overline{\lambda}_{b}$ is a positive constant.
	Similarly we can bound
	$\|\lambda^{*}(\cdot) \nabla_x B_f(x(t))\| \le \overline{\lambda}_{sb}$,
	$\|\tilde{\lambda}(\cdot) \nabla_x B_f(x(t))\| \le \overline{\lambda}_{tb}\;\;\forall\; t \in \mathbb{R}_{\ge t_{0}}$,
	where $\overline{\lambda}_{sb}, \overline{\lambda}_{tb} \in \mathbb{R}_{\ge 0}$
	are positive constants. Furthermore, since all components of the control effort are bounded,
	$\hat{u}(x(t)) \in \mathcal{L}_{\infty}$.
	\qed
\end{pf*}



\section{Stability analysis}
\label{sec:lyap}
Using \eqref{eq:optimalityCondition} we can write the Bellman Error in its unmeasureable form as
\begin{equation}
	\begin{aligned}
		\delta
		 & = r(x,\hat{u})
		+\hat{\lambda} \nabla_x B_f^{T} \hat{F}_{\hat{u}}
		+ \nabla_x \hat{V}^T \hat{F}_{\hat{u}} \\
		 & -r(x,u^*)
		-\lambda^* \nabla_x B_f^{T} F_{u^*}
		-\nabla_x V^{*T} F_{u^*},
	\end{aligned}
\end{equation}
where
\(F_{u^*} \triangleq f+g u^{*}\),
\(\hat{F}_{\hat{u}} \triangleq \hat{f}+g \hat{u}\),
and \(\tilde{F}_{\hat{u}} \triangleq \tilde{f}+g \tilde{u}\).
Substituting \(\hat{u}\) and \(u^{*}\) from \eqref{eq:finalControlLaw} and
\eqref{eq:optimalControlLaw} we have
\begin{equation}
	\begin{aligned}
		 & \delta =
		-\omega^{T}\tilde{W}_c
		+ \frac{1}{2}\tilde{W}_a^{T}R_s\tilde{W}_a
		+\hat{\lambda} \nabla_x B_f(x)^T Y(x) \hat{\theta}                                                                         \\
		 & + \lambda^{*} \hat{W}_a^{T} \nabla_{x} \phi(x) R_g(x)\nabla_x B_f(x)
		+ W^T \nabla_{x} \phi(x) Y(x) \tilde{\theta}                                                                               \\
		 & +(-\frac{1}{2} \tilde{\lambda}^2 - \lambda^{*2}+ \tilde{\lambda} \lambda^{*})\nabla_x B_f(x)^{T}R_{g}(x)\nabla_x B_f(x)
		+ \Delta_{\epsilon},
	\end{aligned}
\end{equation}
where \(\Delta_{\epsilon}\) consists of terms involving the neural network
reconstruction error \(\epsilon\). We can similarly write the BE evaluated at the
extrapolation points \(\{x_{i} = X_{i}(x(t),t)\}_{i=1}^{N}\) as
\begin{equation}
	\begin{aligned}
		 & \delta_{i} =
		-\omega_{i}^{T}\tilde{W}_c
		+ \frac{1}{2}\tilde{W}_a^{T}R_{si}\tilde{W}_a
		+\hat{\lambda} \nabla_x B_f(x_i)^T Y(x_i) \hat{\theta}                                                                                 \\
		 & + \lambda^{*} \hat{W}_a^{T} \nabla_{x} \phi(x_{i}) R_g(x_{i})\nabla_x B_f(x_{i})
		+ W^T \nabla_{x} \phi(x_{i}) Y(x_{i}) \tilde{\theta}                                                                                   \\
		 & +(-\frac{1}{2} \tilde{\lambda}^2 - \lambda^{*2}+ \tilde{\lambda} \lambda^{*})\nabla_x B_f(x_{i})^{T}R_{g}(x_{i})\nabla_x B_f(x_{i})
		+ \Delta_{\epsilon i},
	\end{aligned}
\end{equation}
where \(R_{si} \triangleq R_{s}(x_{i})\) and \(\Delta_{\epsilon i}\) is defined in a
similar fashion to
\(\Delta_{\epsilon}\) with the terms containing state \(x\) replaced by similar
terms involving \(x_{i}\). The errors \(\Delta_{\epsilon}\) and \(\Delta_{\epsilon
	i}\) are uniformly bounded over the domain \(\mathcal{D}\), with the bounds decreasing upon
decreasing \(\|\nabla_{x}\epsilon\|\).

We now define a constant matrix $S$
(cf. \cite{cohen2023Automatica}) containing coefficients of the cross-terms
appearing in the subsequent Lyapunov analysis as
\begin{equation}
	\label{eq:SDefinition}
	\begin{aligned}
		S \triangleq
		\begin{bmatrix}
			\frac{\eta_{c2} \underline{c}}{4} & -\frac{\varphi_{ca}}{2}        & -\frac{\varphi_{ci}}{2} \\
			-\frac{\varphi_{ca}}{2}           & \frac{\eta_{a1}+ \eta_{a2}}{4} & 0                       \\
			-\frac{\varphi_{ci}}{2}           & 0                              & \frac{k_{\theta}}{4}
		\end{bmatrix},
	\end{aligned}
\end{equation}
where
$
	\underline{c} \triangleq \frac{c_{2}}{2} + \frac{\beta}{2 \eta_{c2} \overline{\Gamma}}
$,
$
	\varphi_{ci} \triangleq \frac{\eta_{c1} + \eta_{c2}}{2\sqrt{\nu \underline{\Gamma}}}
	\overline{\|\overline{W}\nabla_{x} \phi Y\|}
$, and
$
	\varphi_{ca} \triangleq \eta_{a1} + \frac{(\eta_{c1} + \eta_{c2})}{4 \sqrt{\nu \underline{\Gamma}}}
	\overline{\|R_{s}\overline{W}\|}
$.
Additionally, we define the positive constant
\begin{equation}
	\begin{aligned}
		\iota & \triangleq
		\overline{\|(W^T \nabla_{x} \phi + \nabla_{x}\epsilon^T)R_g (\nabla_{x}\epsilon + \overline{\lambda}_{tb})\|} \\&
		+\frac{\iota_{c}^{2}}{2 \eta_{c2} \underline{c}}
		+\frac{\iota_{a}^{2}}{2 (\eta_{a1} + \eta_{a2})},
	\end{aligned}
\end{equation}
where
$
	\iota_{c} \triangleq \frac{\eta_{c1} + \eta_{c2}}{2\sqrt{\nu \underline{\Gamma}}} \iota_{\delta}
$,
$\iota_{\delta} \triangleq
	\overline{\|\overline{W} \nabla_{x} \phi R_{g} \overline{\lambda}_{sb}\|}
	+ \frac{5}{2}\overline{\|R_g \overline{\lambda}_{sb}^{2}\|}
	+\overline{\|\overline{\lambda}_{b} Y \overline{\theta}_{h}\|}
	+\frac{1}{2}\overline{\|R_{s}\overline{W}_{at}^2\|}
	+ \overline{\|\Delta_{\epsilon}\|}$, and
$
	\iota_{a} \triangleq
	 \eta_{a2}\overline{W}
	+ \frac{\eta_{c1} + \eta_{c2}}{4\sqrt{\nu \underline{\Gamma}}}
	+ \overline{\|R_{g}(W^{T}\nabla_{x} \phi + \nabla_{x}\epsilon^{T})\|}
	+ \overline{\|R_s^T \overline{W} \overline{W}\|}
$ are positive constants.
It can be shown that there exists a class \(\mathcal{K}\) function
\(\alpha_{3}: \mathbb{R} \rightarrow \mathbb{R}\) such that
$\alpha_{3}(\|\zeta\|) \le
	Q(x)
	+ \frac{\eta_{c2} \underline{c}}{4} \|\tilde{W}_c\|^2
	+ \frac{\eta_{a1}+\eta_{a2}}{4} \|\tilde{W}_a\|^2
	+ \frac{k_{\theta}}{4} \|\tilde{\theta}\|^2,
$
\textcolor{hcolor}{where $\zeta \triangleq [x^{T}, \tilde{W}_c^{T}, \tilde{W}_a^{T}, \tilde{\theta}^{T}]^{T}$
is the augmented state vector.}

We now state the following theorem which demonstrates the stability guarantees of the proposed ACIL algorithm.
\begin{thm}[\textcolor{hcolor}{Practical Stability}]
	\label{thm:stability}
	Under the update laws for actor, critic, identifier and Lagrange multiplier
	discussed in Section \ref{sec:acil}, and provided the sufficient condition
	$S \succcurlyeq 0$, defined in \eqref{eq:SDefinition} holds
	\textcolor{hcolor}{(i.e., $S$ is positive semi-definite)}, the errors in the actor parameter ($\tilde{W}_{a}$), critic
	parameter ($\tilde{W}_{c}$), identifier parameter ($\tilde{\theta}$) for the
	system in \eqref{eq:plantDynamics} are uniformly ultimately bounded (UUB).
	Additionally, the state $x$ is bounded with the set $\mathcal{C}$ being forward
	invariant for the system in \eqref{eq:plantDynamics}.
\end{thm}
\vspace*{-0.5cm}
\begin{pf*}{Proof}
	We consider the positive-definite candidate Lyapunov function
	\(V_{L}(\zeta,t): \mathbb{R}^{n+2b+p} \times \mathbb{R}_{>t_{0}} \rightarrow \mathbb{R}\) defined as
	\begin{equation}
		V_{L}(\zeta,t) = V^{*}(x)
		+ \frac{1}{2}\tilde{W}_c^{T}\Gamma(t)^{-1}\tilde{W}_c
		+ \frac{1}{2}\tilde{W}_a^{T}\tilde{W}_a
		+ V_{\theta}(\tilde{\theta},t).
	\end{equation}
	We observe that there exist two class
	\(\mathcal{K}\) functions \(\alpha_{1}(\cdot)\) and \(\alpha_{2}(\cdot)\) \cite[Lemma 4.3]{khalil2002nonlinear} such that
	$\alpha_{1}(\|\zeta\|) \le
		V_{L}(\zeta,t) \le
		\alpha_{2}(\|\zeta\|) \;\;\forall\; t \in [t_{0},\infty).$
	The time derivative of the candidate Lyapunov function along the
	trajectories of the closed loop system is
	\begin{equation}
		\begin{aligned}
			 & \dot{V}_{L}(\zeta,t) =
			\nabla_x V^{*T} (f(x) + g(x) \hat{u})
			- \tilde{W}_c^T \Gamma^{-1} \dot{\hat{W}}_c                                     \\
			 & -\frac{1}{2}\tilde{W}_c^{T} \Gamma^{-1} \dot{\Gamma} \Gamma^{-1} \tilde{W}_c
			-\tilde{W}_a^T \dot{\hat{W}}_a
			+ \dot{V}_{\theta}.
		\end{aligned}
	\end{equation}
	Substituting the update laws for the actor, critic, least-squares gain matrix and
	identifier from Section \ref{sec:acil}, we obtain the bound
	\begin{equation}
		\dot{V}_{L}(\zeta,t) \le - \alpha_{3}(\|\zeta\|) - \zeta_{1}^{T}S \zeta_{1} + \iota,
	\end{equation}
	where
	$\zeta_{1} \triangleq [\tilde{W}_{c}^{T},\tilde{W}_{a}^{T}, \tilde{\theta}^{T}]^{T}$.
	Under the sufficient condition $S \succcurlyeq 0$, we can obtain the bound
	$\dot{V}_{L}(\zeta,t) \le - \alpha_{3}(\|\zeta\|)  + \iota.$
	We observe that $\dot{V}_{L}(\cdot)$ is negative outside the compact set
	$\Omega_{s} \triangleq \{\zeta \in \mathbb{R}^{n+2b+p}  :\|\zeta\| \le \alpha_{3}^{-1}(\iota)\}$. Using \cite[Theorem 4.18]{khalil2002nonlinear}
	we conclude that the augmented state $\zeta$ is uniformly ultimately bounded.
	While the bound on state $x$ is established by the UUB result mentioned above,
	we obtain a much stronger result from Theorem \ref{thm:safety}, i.e.,
	$x(t) \in \mathcal{C} \;\;\forall\; t \in \mathbb{R}_{\ge t_{0}}$. \qed
\end{pf*}
\begin{rmark}
	The ultimate bound established in Theorem \ref{thm:stability} depends on the constant $\iota$. One
	can reduce the value of the constant by increasing the neurons in the neural
	network estimating the value function and increasing the gains of
	the actor, critic, identifier and the estimated Lagrange multiplier.
\end{rmark}





\section{Simulation Results}
\label{sec:results}
\subsection{Wing rock stabilization for a Delta-wing aircraft}
To demonstrate the efficacy of the proposed ACIL algorithm, we consider the
wing rock stabilization problem for a Delta-wing aircraft \cite{hsu1985theory}, which is a highly nonlinear phenomenon.
We consider the constrained optimal control for the dynamics
\textcolor{hcolor}{
	\begin{equation}
		\begin{aligned}
			\dot{\varphi} & = p, \\
			\dot{p}       & =
			\theta_{1}\varphi
			+ \theta_{2}p
			+ \theta_{3} |\varphi| p
			+\theta_{4} |p|p
			+ \theta_{5}\varphi^{3}
			+ \theta_{6} u,
		\end{aligned}
	\end{equation}
}
where $\varphi(t) \in \mathbb{R}$ is the aircraft roll angle (in $\unit{rad}$),
$p(t) \in \mathbb{R}$ is the roll rate (in $\unit{rad.s^{-1}}$), and
$u(t) \in \mathbb{R}$ is the differential aileron input (in $\unit{rad}$). The
values of the parameters considered are
$
	\theta_1=-0.018,
	\theta_2=0.015,
	\theta_3=-0.062,
	\theta_4=0.009,
	\theta_5=0.021,
	\theta_6=0.75,
$
where only $\theta_{6}$ is known a-priori. The instantaneous state cost function
is considered as
$Q(x) = x^{T}x$, where $x \triangleq [\varphi,p]^{T}$ is the state vector and
$R = 1$. We impose a user-defined constraint
$\|x(t)\| < 2 \;\;\forall\; t \in \mathbb{R}_{\ge 0}$, with the corresponding
candidate BLF as $B_{f}(x) = (\frac{4}{4-x^{T}x} - 1)^{2}$.
The basis function was chosen as
$\phi(x) = [\varphi^{2},p^{2},\varphi p, \varphi^{3} p]^{T}$ and the actor and
critic parameters were initialized as
$\hat{W}_{a}(0) = \hat{W}_{c}(0) = [10,10,10,0]^{T}$ with $\Gamma(0) = 10\mathbb{I}_{4}$. We considered the gains
$\eta_{c1} = 0.1$, $\eta_{c2} = 1$, $\eta_{a1} = 0.1$, $\eta_{a2} = 1$,
$\nu = 5$, and $\beta = 0.01$. The softplus gain was taken to be $k = 0.02$ and
the constant $k_{sb}$ was set to $0.2$.

\begin{figure}[htbp]
	\centering
	\begin{subfigure}[t]{0.8\linewidth}
		\centering
		\includegraphics[width=0.9\textwidth]{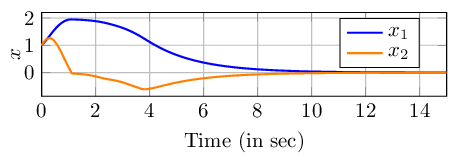}
		\vspace*{-0.4cm}
		\caption{State Trajectory}
		\vspace*{0.35cm}
		\label{subfig:nonlin_x}
	\end{subfigure}
	\hfill
	\begin{subfigure}[t]{0.8\linewidth}
		\centering
		\includegraphics[width=0.9\textwidth]{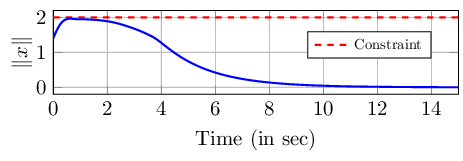}
		\vspace*{-0.4cm}
		\caption{Constraint satisfaction}
		\vspace*{0.35cm}
		\label{subfig:nonlin_xnorm}
	\end{subfigure}
	\hfill
	\begin{subfigure}[t]{0.8\linewidth}
		\centering
		\includegraphics[width=0.9\textwidth]{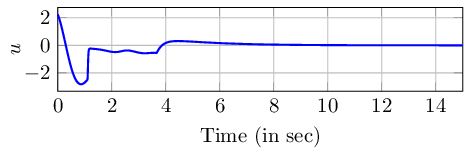}
		\vspace*{-0.4cm}
		\caption{Control Effort}
		\vspace*{0.35cm}
		\label{subfig:nonlin_u}
	\end{subfigure}
	\hfill
	\begin{subfigure}[t]{0.8\linewidth}
		\centering
		\includegraphics[width=0.9\textwidth]{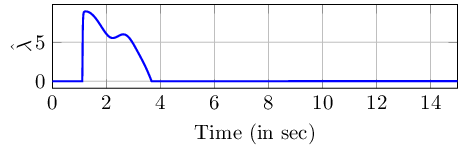}
		\vspace*{-0.4cm}
		\caption{Lagrange multiplier estimate}
		\vspace*{0.35cm}
		\label{subfig:nonlin_lambda}
	\end{subfigure}
	\hfill
	\begin{subfigure}[t]{0.8\linewidth}
		\centering
		\includegraphics[width=0.9\textwidth]{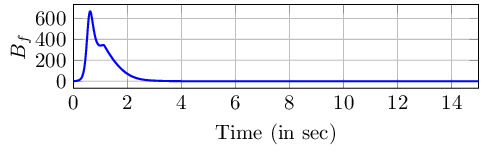}
		\vspace*{-0.4cm}
		\caption{Barrier value}
		\vspace*{0.4cm}
		\label{subfig:nonlin_barrier}
	\end{subfigure}
	\vspace*{-0.5cm}
	\caption{Relevant plots for the delta wing system}
	\label{fig:nonlinear}
\end{figure}

Fig. \ref{fig:nonlinear} shows the simulation plots for the delta-wing system.
Fig. \ref{subfig:nonlin_x} shows the state trajectory for the delta-wing
system under the proposed ACIL algorithm. The constraint satisfaction by the
proposed algorithm is shown in Fig. \ref{subfig:nonlin_xnorm}.
We see that the norm of the state is
always below the prescribed limit of two (red dotted line).
Additionally, the states converge to
zero within 11 seconds of the simulation. Fig. \ref{subfig:nonlin_u},
\ref{subfig:nonlin_lambda} and \ref{subfig:nonlin_barrier} show the control effort,
estimated Lagrange multiplier and barrier function plots respectively.
We observe that the proposed on-policy ACIL approach successfully achieves the regulation
objectives while satisfying the user-defined constraint on state.
\begin{table*}[htbp]
	\caption{Cost comparision under different initial conditions \label{tab:massive}}
	\centering
	\resizebox{0.8\linewidth}{!}{\begin{tabular}{|c|c|c|c|c|c|c|}
			\hline
			\multirow{2}{*}{Method}
			 & \multicolumn{3}{c|}{Mobile robot system}

			 & \multicolumn{3}{c|}{Delta wing system}                                                        \\
			\cline{2-7}
			 & $x_{0} = [4,6]^{T}$                      & $x_{0} = [-7.5,4.5]^{T}$ & $x_{0} = [1,-9.2]^{T}$

			 & $x_{0} = [1,0.1]^{T}$                    & $x_{0} = [-1,1]^{T}$     & $x_{0} = [1.9,0.1]^{T}$

			\\
			\hline
			ACIL (proposed)
			 & \textbf{107.521}                         & \textbf{157.371}         & 176.693
			 & \textbf{13.493}                          & \textbf{4.069}           & \textbf{11.270}         \\
			
			\cite{cohen2023Automatica}
			 & 111.201                                  & 158.460                  & \textbf{173.076}

			 & 14.273                                   & 4.074                    & 15.704                  \\
			\hline
			
			ACIL (proposed) with known $\theta$
			 & \textbf{91.107 }                         & \textbf{136.811}         & \textbf{161.481 }

			 & \textbf{4.835}                           & \textbf{4.046}           & \textbf{12.075}         \\
			
			\cite{cohen2023Automatica} with known $\theta$
			 & 98.291                                   & 148.323                  & 164.911
			 & 4.886                                    & 4.048                    & 15.708                  \\
			\hline
			
			SGSA\cite{almubarak2021CDC} (offline with known $\theta$)
			 & 67.892                                   & 99.282                   & 93.444
			 & 2.338                                    & 2.270                    & 8.205                   \\
			\hline
		\end{tabular}}
\end{table*}



\subsection{Mobile robot in a minefield}
To demonstrate the efficacy of the proposed ACIL algorithm on non-convex state
constraints, we
\textcolor{hcolor}{
	consider a mobile robot with the dynamics
	\begin{equation}
		\dot{x} =
		\begin{bmatrix}
			\theta_{1} & \theta_{2} \\
			\theta_{3} & \theta_{4} \\
		\end{bmatrix} x
		+
		\begin{bmatrix}
			1 & 0 \\
			0 & 1 \\
		\end{bmatrix} u,
	\end{equation}
	where
	$x(t),u(t) \in \mathbb{R}^{2}$ and the unknown parameters
	$\theta_{1},\theta_{2},\theta_{3},\theta_{4} \in \mathbb{R} $
	and the true values of the parameters are
	$\theta_{1} = \theta_{2} = \theta_{3} = \theta_{4} = 0$.
}
The objective of the mobile robot is to reach
the origin while avoiding ``mines'' strewn across a ``minefield''. We consider
twelve mines placed randomly inside a circle of radius 10 centered
at origin. The $i^{th}$ obstacle is considered to be a unit circle with center
at $c_{i} \;\;\forall\; i \in \mathbb{N}_{12}$ . To incorporate this complex
constraint on the state, we consider the BLF
$B_{f} = B_{field} + \sum_{i=1}^{12}B_{obs_{i}}$, where
$B_{field} = (\frac{100}{100 - x^{T}x} - 1)^{2}$ is the barrier function for the
field and $B_{obs_{i}} = \frac{1}{\|x-c_{i}\|^{2} - 1} $ is the barrier function
for an individual obstacle.
We choose
$Q(x) = x^{T}x$ and $R = \mathbb{I}_{2}$ for the given
optimal regulation problem.
To learn an optimal control policy, we consider the
basis function $\phi(x) = [x_{1}^{2},x_{1}x_{2},x_{2}^{2}]^{T}$ and the actor-critic
parameters were initialized as $\hat{W}_{a}(0) = \hat{W}_{c}(0) = [2,0,2]^{T}$.
The gains for the actor-critic and the Lagrangian components of the algorithm were taken to be the
same values as that of the delta-wing system.

We compare the proposed method with that of
\cite{cohen2023Automatica}, which is an online RL algorithm. The algorithm in
\cite{cohen2023Automatica} can be considered as the ACIL method with a constant
$\hat{\lambda}$ (Technically, the constant $c_{b}$ used in \cite{cohen2023Automatica}
is related to a constant $\hat{\lambda}$ by the relation
$c_{b} = R^{-1}\hat{\lambda}$). To enable a fair comparision and have
equivalent control effort applied, we set $c_{b} = 0.075$ and keep the gains for
the Actor Critic and Identifier components of \cite{cohen2023Automatica} to be
the same as the gains for the proposed algorithm.  We additionally
compare both these methods with an offline numerical method of safe Galerkin
successive approximation (SGSA) detailed in \cite{almubarak2021CDC}, which
assumes the complete knowledge of the system dynamics, and generates control policies close to the
optimal solution. The SGSA algorithm in \cite{almubarak2021CDC} utilizes a
higher-order zeroing control barrier function to ensure safety of the RL agent.

\begin{figure}[htbp]
	\centerline{\includegraphics[width=0.8\columnwidth]{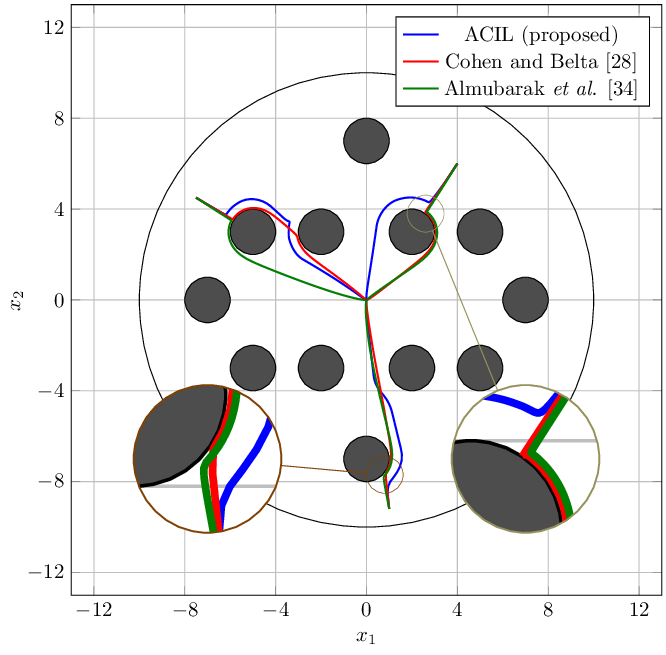}}
	\vspace*{-0.4cm}
	\caption{Visualization of the mobile robot system under different initial
		conditions for different RL algorithms. The blue line-plot is that of ACIL
		(proposed), the red line-plot is that of the algorithm detailed in
		\cite{cohen2023Automatica}, and the green line-plot is that of the offline
		algorithm SGSA detailed in \cite{almubarak2021CDC}.      \label{fig:circle_plot} }
\end{figure}

\begin{figure}[htpb]
	\centering
	\begin{subfigure}[t]{\linewidth}
		\centering
		\includegraphics[width=0.9\columnwidth]{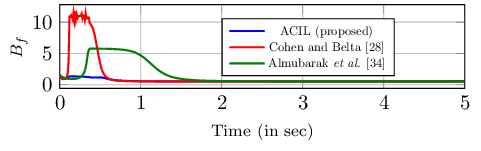}
		\vspace*{-0.4cm}
		\caption{Barrier value}
		\vspace*{0.4cm}
		\label{subfig:lin_aux_barrier}
	\end{subfigure}
	\hfill
	\begin{subfigure}[t]{\linewidth}
		\centering
		\includegraphics[width=0.9\columnwidth]{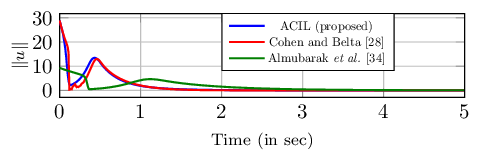}
		\vspace*{-0.4cm}
		\caption{Control effort}
		\label{subfig:comp_u}
	\end{subfigure}
	\caption{Comparision of the proposed method with
		\cite{cohen2023Automatica} and \cite{almubarak2021CDC} for the mobile robot system}
	\label{fig:linear_aux}
\end{figure}

Fig. \ref{fig:circle_plot} shows the visualization of the three algorithms
on the mobile robot system. Fig. \ref{fig:linear_aux} shows the (a) barrier
function and (b) control effort for the three algorithms. We
observe that both ACIL and \cite{cohen2023Automatica} apply equivalent
control effort, but the peak value of the barrier function is less for the proposed
ACIL algorithm. Additionally, we compare the cost inccured by each of these
algorithms for both the mobile robot system and the delta-wing system in Table
\ref{tab:massive}. We also compare the costs for ACIL and that of
\cite{cohen2023Automatica} under the complete knowledge of the drift dynamics.
We observe that in both the systems, the proposed ACIL method incurs lesser
cost than that of \cite{cohen2023Automatica}. Provided the knowledge of the
system dynamics, the proposed method outperforms \cite{cohen2023Automatica}.


\subsection{Effect of varying the softplus gain ($k$)}
We now study the effect of varying the softplus gain $k$ on the case of mobile
robot system. Under different values of $k$ we tabulate the cost incurred in
Table \ref{tab:kVar}.

\begin{figure}[htpb]
	\centering
	\begin{subfigure}[t]{\linewidth}
		\centering
		\includegraphics[width=0.9\columnwidth]{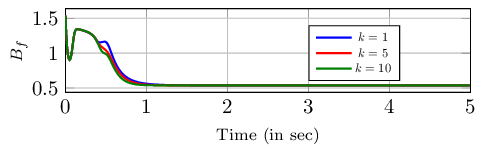}
		\vspace*{-0.4cm}
		\caption{Barrier value}
		\vspace*{0.3cm}
	\end{subfigure}
	\hfill
	\begin{subfigure}[t]{\linewidth}
		\centering
		\includegraphics[width=0.9\columnwidth]{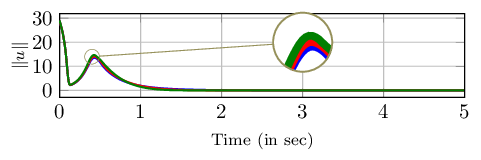}
		\vspace*{-0.4cm}
		\caption{Control effort}
	\end{subfigure}
	\caption{Effect of varying the gain $k$ of the softplus function}
	\label{fig:kVariation}
\end{figure}

\begin{table}[htbp]
	\caption{Accrued cost under different values of softplus gain $k$ for the mobile
		robot system. \label{tab:kVar}}
	\centering
	\resizebox{0.8\columnwidth}{!}{\begin{tabular}{|c|c|c|c|c|}
			\hline
			$k = 0.02$ & $k = 0.1$ & $k = 1$ & $k = 5$ & $k = 10$ \\
			\hline
			107.477    & 107.506   & 108.078 & 112.149 & 117.121  \\
			\hline
		\end{tabular}}
\end{table}
Fig. \ref{fig:kVariation} shows the barrier function and the norm of control action.
We observe that as the value of $k$
increases, the cost performance deteriorates. Additionally, the control effort
required increases with the increase in $k$. However, with the increase of $k$,
the peak value of the BLF near the obstacles decreases.



\section{Conclusion}
We develop an on-policy reinforcement learning algorithm for optimal
control of a class of uncertain nonlinear systems utilizing an online
estimation of Lagrange multipliers. We extend the ACI approach from
\cite{bhasin2013Automatica,kamalapurkar2016Automatica} to learn optimal control policies in a safe
fashion. We prove the safety and stability guarantees of the proposed
algorithm via a Lyapunov analysis. We subsequently demonstrate the efficacy of
the ACIL method on two systems in simulation. Additionally, we show that the
proposed method outperforms a similar online algorithm in the literature.
Future work will seek to include actuation constraints in the formulation to
handle state and actuation constraints in a combined way, \textcolor{hcolor}{and
	also relax the restrictive assumption of a known $g(x)$, considered in
	this work. }

\appendix
\section{Proof of Lemma \ref{lem:linearityOfSoftplus}}\label{append:proof}
The first and second derivatives of $f_{\sigma}$ on $(0,\infty)$ are
\begin{equation}
	\label{eq:firstSecondDerivative}
	\begin{aligned}
		f_{\sigma}'(z)  & = \sigma\Big(\frac{c}{z}\Big) - \frac{c}{z}\sigma'\Big(\frac{c}{z}\Big), \\
		f_{\sigma}''(z) & =  \frac{c^{2}}{z^{3}}\sigma''\Big(\frac{c}{z}\Big),
	\end{aligned}
\end{equation}
respectively. Using Properties P1-P3, we show that $0 \le \sigma'(x) \le 1$ and
$0 \le \sigma''(x) \;\;\forall\;x \in \mathbb{R}$. Consequently, it can be shown
that $f_{\sigma}'(z) \ge 0$ and $f_{\sigma}''(z) \ge 0 \;\;\forall\; z \in (0,\infty)$.
In other words $f_{\sigma}(z)$ and $f_{\sigma}'(z)$ are monotonically increasing functions in the domain
$(0,\infty)$.
Using the mean-value theorem, for any interval $(h,z) \subset (0,\infty)$,
$\exists \; \upsilon \in (h,z)$ such that
\begin{equation}
	\frac{f_{\sigma }(z) - f_{\sigma }(h)}{z-h} = f'_{\sigma }(\upsilon) \leq \sup_{x \in (0,\infty)} |f_{\sigma }'(x)|.
\end{equation}
If $h\to 0_{+}$, we can evaluate the limit
$
	\lim_{h\rightarrow 0+} f_{\sigma}(h) = \lim_{h\rightarrow 0+} \sigma (\frac{c}{h}) h = c,
$
via the L'H\^opital's rule. We can thus write
$\frac{f_{\sigma}(z)-c}{z} \le \sup_{x \in (0,\infty)} | f_{\sigma}'(x) | \;\;\forall\; z \in (0,\infty).$
Since $f_{\sigma}'$ is monotonically increasing, the supremum of $f_{\sigma}'$
exists at infinity. Thus computing the said limit, we have
$
	\frac{f_{\sigma}(z)-c}{z} \le \sigma(0)
	\;\;\forall\; z \in (0,\infty),
$
which on rearrangement, completes the proof.

\bibliographystyle{ieeetr}        
\bibliography{uni}           

\begin{thebibliography}{10}

\bibitem{bhatnagar2009Automatica}
S.~Bhatnagar, R.~S. Sutton, M.~Ghavamzadeh, and M.~Lee, ``Natural actor--critic
  algorithms,'' {\em Automatica}, vol.~45, no.~11, pp.~2471--2482, 2009.

\bibitem{sutton2018reinforcement}
R.~S. Sutton and A.~G. Barto, {\em Reinforcement learning: An introduction}.
\newblock MIT press, 2018.

\bibitem{al2008TSMC}
A.~Al-Tamimi, F.~L. Lewis, and M.~Abu-Khalaf, ``Discrete-time nonlinear hjb
  solution using approximate dynamic programming: Convergence proof,'' {\em
  IEEE Transactions on Systems, Man, and Cybernetics, Part B (Cybernetics)},
  vol.~38, no.~4, pp.~943--949, 2008.

\bibitem{vamvoudakis2010Automatica}
K.~G. Vamvoudakis and F.~L. Lewis, ``Online actor--critic algorithm to solve
  the continuous-time infinite horizon optimal control problem,'' {\em
  Automatica}, vol.~46, no.~5, pp.~878--888, 2010.

\bibitem{bhasin2013Automatica}
S.~Bhasin, R.~Kamalapurkar, M.~Johnson, K.~G. Vamvoudakis, F.~L. Lewis, and
  W.~E. Dixon, ``A novel actor--critic--identifier architecture for approximate
  optimal control of uncertain nonlinear systems,'' {\em Automatica}, vol.~49,
  no.~1, pp.~82--92, 2013.

\bibitem{pang2022TAC}
B.~Pang and Z.-P. Jiang, ``Reinforcement learning for adaptive optimal
  stationary control of linear stochastic systems,'' {\em IEEE Transactions on
  Automatic Control}, vol.~68, no.~4, pp.~2383--2390, 2022.

\bibitem{ames2016TAC}
A.~D. Ames, X.~Xu, J.~W. Grizzle, and P.~Tabuada, ``Control barrier function
  based quadratic programs for safety critical systems,'' {\em IEEE
  Transactions on Automatic Control}, vol.~62, no.~8, pp.~3861--3876, 2016.

\bibitem{ames2019ECC}
A.~D. Ames, S.~Coogan, M.~Egerstedt, G.~Notomista, K.~Sreenath, and P.~Tabuada,
  ``Control barrier functions: Theory and applications,'' in {\em 2019 18th
  European control conference (ECC)}, pp.~3420--3431, IEEE, 2019.

\bibitem{koller2018CDC}
T.~Koller, F.~Berkenkamp, M.~Turchetta, and A.~Krause, ``Learning-based model
  predictive control for safe exploration and reinforcement learning,'' {\em
  Proceedings of the IEEE Conference on Decision and Control}, pp.~6059--6066,
  2018.

\bibitem{zanon2020TAC}
M.~Zanon and S.~Gros, ``Safe reinforcement learning using robust mpc,'' {\em
  IEEE Transactions on Automatic Control}, vol.~66, no.~8, pp.~3638--3652,
  2020.

\bibitem{berkenkamp2017NIPS}
F.~Berkenkamp, M.~Turchetta, A.~Schoellig, and A.~Krause, ``Safe model-based
  reinforcement learning with stability guarantees,'' {\em Advances in neural
  information processing systems}, vol.~30, 2017.

\bibitem{cohen2021TL}
M.~H. Cohen and C.~Belta, ``Model-based reinforcement learning for approximate
  optimal control with temporal logic specifications,'' in {\em Proceedings of
  the 24th International Conference on Hybrid Systems: Computation and
  Control}, pp.~1--11, 2021.

\bibitem{alshiekh2018safe}
M.~Alshiekh, R.~Bloem, R.~Ehlers, B.~K{\"o}nighofer, S.~Niekum, and U.~Topcu,
  ``Safe reinforcement learning via shielding,'' in {\em Proceedings of the
  AAAI Conference on Artificial Intelligence}, vol.~32, 2018.

\bibitem{fisac2018general}
J.~F. Fisac, A.~K. Akametalu, M.~N. Zeilinger, S.~Kaynama, J.~Gillula, and
  C.~J. Tomlin, ``A general safety framework for learning-based control in
  uncertain robotic systems,'' {\em IEEE Transactions on Automatic Control},
  vol.~64, no.~7, pp.~2737--2752, 2018.

\bibitem{cheng2019AAAI}
R.~Cheng, G.~Orosz, R.~M. Murray, and J.~W. Burdick, ``End-to-end safe
  reinforcement learning through barrier functions for safety-critical
  continuous control tasks,'' {\em 33rd AAAI Conference on Artificial
  Intelligence, AAAI 2019, 31st Innovative Applications of Artificial
  Intelligence Conference, IAAI 2019 and the 9th AAAI Symposium on Educational
  Advances in Artificial Intelligence, EAAI 2019}, pp.~3387--3395, 2019.

\bibitem{blanchini1999Automatica}
F.~Blanchini, ``Set invariance in control,'' {\em Automatica}, vol.~35, no.~11,
  pp.~1747--1767, 1999.

\bibitem{tee2009Automatica}
K.~P. Tee, S.~S. Ge, and E.~H. Tay, ``Barrier lyapunov functions for the
  control of output-constrained nonlinear systems,'' {\em Automatica}, vol.~45,
  no.~4, pp.~918--927, 2009.

\bibitem{hager2018SIAM}
W.~W. Hager, J.~Liu, S.~Mohapatra, A.~V. Rao, and X.-S. Wang, ``Convergence
  rate for a gauss collocation method applied to constrained optimal control,''
  {\em SIAM Journal on Control and Optimization}, vol.~56, no.~2,
  pp.~1386--1411, 2018.

\bibitem{hager2019COA}
W.~W. Hager, H.~Hou, S.~Mohapatra, A.~V. Rao, and X.-S. Wang, ``Convergence
  rate for a radau hp collocation method applied to constrained optimal
  control,'' {\em Computational Optimization and Applications}, vol.~74, no.~1,
  pp.~275--314, 2019.

\bibitem{rao2009survey}
A.~V. Rao, ``A survey of numerical methods for optimal control,'' {\em Advances
  in the Astronautical Sciences}, vol.~135, no.~1, pp.~497--528, 2009.

\bibitem{vamvoudakis2017SysConLet}
K.~G. Vamvoudakis, ``Q-learning for continuous-time linear systems: A
  model-free infinite horizon optimal control approach,'' {\em Systems \&
  Control Letters}, vol.~100, pp.~14--20, 2017.

\bibitem{kamalapurkar2016Automatica}
R.~Kamalapurkar, J.~A. Rosenfeld, and W.~E. Dixon, ``Efficient model-based
  reinforcement learning for approximate online optimal control,'' {\em
  Automatica}, vol.~74, pp.~247--258, 2016.

\bibitem{yang2019ACC}
Y.~Yang, Y.~Yin, W.~He, K.~G. Vamvoudakis, H.~Modares, and D.~C. Wunsch,
  ``Safety-aware reinforcement learning framework with an actor-critic-barrier
  structure,'' in {\em 2019 American Control Conference (ACC)}, pp.~2352--2358,
  IEEE, 2019.

\bibitem{mahmud2021ACC}
S.~N. Mahmud, K.~Hareland, S.~A. Nivison, Z.~I. Bell, and R.~Kamalapurkar, ``A
  safety aware model-based reinforcement learning framework for systems with
  uncertainties,'' in {\em 2021 American Control Conference (ACC)},
  pp.~1979--1984, IEEE, 2021.

\bibitem{greene2020LCSS}
M.~L. Greene, P.~Deptula, S.~Nivison, and W.~E. Dixon, ``Sparse learning-based
  approximate dynamic programming with barrier constraints,'' {\em IEEE Control
  Systems Letters}, vol.~4, no.~3, pp.~743--748, 2020.

\bibitem{marvi2021IJRNC}
Z.~Marvi and B.~Kiumarsi, ``Safe reinforcement learning: A control barrier
  function optimization approach,'' {\em International Journal of Robust and
  Nonlinear Control}, vol.~31, no.~6, pp.~1923--1940, 2021.

\bibitem{cohen2020CDC}
M.~H. Cohen and C.~Belta, ``Approximate optimal control for safety-critical
  systems with control barrier functions,'' in {\em 2020 59th IEEE Conference
  on Decision and Control (CDC)}, pp.~2062--2067, IEEE, 2020.

\bibitem{cohen2023Automatica}
M.~H. Cohen and C.~Belta, ``Safe exploration in model-based reinforcement
  learning using control barrier functions,'' {\em Automatica}, vol.~147,
  p.~110684, 2023.

\bibitem{bandyopadhyay2023CDC}
S.~Bandyopadhyay and S.~Bhasin, ``Safe q-learning for continuous-time linear
  systems,'' in {\em 2023 62nd IEEE Conference on Decision and Control (CDC)},
  pp.~241--246, IEEE, 2023.

\bibitem{isaly2021ACC}
A.~Isaly, O.~S. Patil, R.~G. Sanfelice, and W.~E. Dixon, ``Adaptive safety with
  multiple barrier functions using integral concurrent learning,'' in {\em 2021
  American Control Conference (ACC)}, pp.~3719--3724, IEEE, 2021.

\bibitem{kokolakis2022TNNLS}
N.-M.~T. Kokolakis and K.~G. Vamvoudakis, ``Safety-aware pursuit-evasion games
  in unknown environments using gaussian processes and finite-time convergent
  reinforcement learning,'' {\em IEEE Transactions on Neural Networks and
  Learning Systems}, pp.~3130--3143, 2022.

\bibitem{peng2023TSMC}
C.~Peng, X.~Liu, and J.~Ma, ``Design of safe optimal guidance with obstacle
  avoidance using control barrier function-based actor--critic reinforcement
  learning,'' {\em IEEE Transactions on Systems, Man, and Cybernetics:
  Systems}, pp.~6861--6873, 2023.

\bibitem{cohen2023LCSS}
M.~H. Cohen, P.~Ong, G.~Bahati, and A.~D. Ames, ``Characterizing smooth safety
  filters via the implicit function theorem,'' {\em IEEE Control Systems
  Letters}, 2023.

\bibitem{almubarak2021CDC}
H.~Almubarak, E.~A. Theodorou, and N.~Sadegh, ``Hjb based optimal safe control
  using control barrier functions,'' in {\em 2021 60th IEEE Conference on
  Decision and Control (CDC)}, pp.~6829--6834, IEEE, 2021.

\bibitem{boyd2004convex}
S.~Boyd and L.~Vandenberghe, {\em Convex optimization}.
\newblock Cambridge university press, 2004.

\bibitem{lewis2012optimal}
F.~L. Lewis, D.~Vrabie, and V.~L. Syrmos, {\em Optimal control}.
\newblock John Wiley \& Sons, third~ed., 2012.

\bibitem{khalil2002nonlinear}
H.~K. Khalil, ``Nonlinear systems,'' {\em Prentice Hall, Upper Saddle River},
  2002.

\bibitem{kreinovich1991NN}
V.~Y. Kreinovich, ``Arbitrary nonlinearity is sufficient to represent all
  functions by neural networks: a theorem,'' {\em Neural networks}, vol.~4,
  no.~3, pp.~381--383, 1991.

\bibitem{nair2010rectified}
V.~Nair and G.~E. Hinton, ``Rectified linear units improve restricted boltzmann
  machines,'' in {\em Proceedings of the 27th International Conference on
  International Conference on Machine Learning}, ICML'10, (Madison, WI, USA),
  p.~807–814, Omnipress, 2010.

\bibitem{lavretsky2013robust}
E.~Lavretsky and K.~A. Wise, ``Robust adaptive control,'' in {\em Robust and
  adaptive control}, pp.~317--353, Springer, 2013.

\bibitem{roy2017TAC}
S.~B. Roy, S.~Bhasin, and I.~N. Kar, ``Combined mrac for unknown mimo lti
  systems with parameter convergence,'' {\em IEEE Transactions on Automatic
  Control}, vol.~63, no.~1, pp.~283--290, 2017.

\bibitem{parikh2019integral}
A.~Parikh, R.~Kamalapurkar, and W.~E. Dixon, ``Integral concurrent learning:
  Adaptive control with parameter convergence using finite excitation,'' {\em
  International Journal of Adaptive Control and Signal Processing}, vol.~33,
  no.~12, pp.~1775--1787, 2019.

\bibitem{hsu1985theory}
C.-H. Hsu and C.~E. Lan, ``Theory of wing rock,'' {\em Journal of Aircraft},
  vol.~22, no.~10, pp.~920--924, 1985.

\end{thebibliography}



\end{document}